\documentclass[reprint,aps,prx,superscriptaddress,longbibliography]{revtex4-1}

\usepackage{braket}
\usepackage{amsmath,amsthm,mathrsfs,bm,bbm}	
\usepackage{empheq}
\usepackage[citecolor=black,linkcolor=black]{hyperref} 

\begin{document}

\title{Generalized theory of pseudomodes for exact descriptions of non-Markovian quantum processes}

\author{Graeme Pleasance}
\email{gpleasance1@gmail.com}
\affiliation{Quantum Research Group, School of Chemistry and Physics, University of KwaZulu-Natal, Durban, 4001, South Africa}

\author{Barry M. Garraway}
\affiliation{Department of Physics and Astronomy, University of Sussex, Falmer, Brighton, BN1 9QH, United Kingdom}

\author{Francesco Petruccione}
\affiliation{Quantum Research Group, School of Chemistry and Physics, University of KwaZulu-Natal, Durban, 4001, South Africa}
\affiliation{National Institute for Theoretical Physics (NITheP), KwaZulu-Natal, South Africa}
\affiliation{School of Electrical Engineering, KAIST, Daejeon, 34141, Republic of Korea}

\date{\today}

\begin{abstract}

We develop an exact framework for describing the non-Markovian dynamics of an open quantum system interacting with an environment modeled by a generalized spectral density function. The approach relies on mapping the initial system onto an auxiliary configuration, comprising the original open system coupled to a small number of discrete modes, which in turn are each coupled to an independent Markovian reservoir. Based on the connection between the discrete modes and the poles of the spectral density function, we show how expanding the system using the discrete modes allows for the full inclusion non-Markovian effects within an enlarged open system whose dynamics is governed by an exact Lindblad master equation. Initially we apply this result to obtain a generalization of the pseudomode method [B. M. Garraway, Phys. Rev. A \textbf{55}, 2290 (1997)] in cases where the spectral density function has a Lorentzian structure. For many other types of spectral density function, we extend our proof to show that an open system dynamics may be modeled physically using discrete modes which admit a non-Hermitian coupling to the system, and for such cases determine the equivalent master equation to no longer be of Lindblad form. For applications involving two discrete modes, we demonstrate how to convert between pathological and Lindblad forms of the master equation using the techniques of the pseudomode method.

\end{abstract}

\maketitle

\section{Introduction}\label{sec:1}

The theory of open quantum systems, which concerns the interaction between a quantum system of interest and a large macroscopic reservoir or heat bath \cite{BreuerTOQS2002,Gardiner2005}, plays a fundamental role in several applications of quantum physics, ranging from quantum information \cite{Nielson2010}, quantum technologies, and decoherence \cite{Schlosshauer2019,Zurek1991}, through to quantum optics \cite{Carmichael1993}, condensed matter \cite{Leggett1987}, and quantum thermodynamics \cite{Vinjanampathy2016}. In many applications a standard approach is to model the effect of the environment in terms of a Markovian master equation, whose general validity requires the environmental noise (as measured through the reservoir correlation function) to be correlated over a much shorter time interval than the characteristic decay time of the open system. This condition is known to be well satisfied in quantum optical and mesoscopic systems \cite{Tudela2013,*Brandes2005} where the reservoir coupling constants vary slowly with frequency, and the coupling to the system of interest is typically weak.

For many other situations, however, especially those involving environments that are structured\textemdash i.e., with long correlation times and frequency-dependent coupling constants\textemdash the assumption of a large separation of timescales between the system and environment no longer applies, and for these cases the resulting dynamics is non-Markovian \cite{Breuer2016,DeVega2017,Li2018}. In recent years there has been renewed attention paid to non-Markovian open quantum systems, not only out of fundamental interest but also due to the growing number of practical applications. From one perspective, a wide variety of experimental platforms, including atom-cavity and trapped-ion systems \cite{Lemmer2018}, solid-state devices \cite{Ribeiro2015}, and photonic-band gap materials \cite{Hoeppe2012}, have been shown to feature regimes where non-Markovian and strong-coupling effects play an significant role in the description of the dynamics. At the same time, the increasing ability to coherently control the non-Markovian dynamics of quantum systems through, e.g., the use of reservoir engineering techniques \cite{Haase2018,Wang2018,Peng2018,Liu2011,Ho2019,Liu2018}, has provided new avenues to explore how certain types of environmental noise might be useful for the implementation of quantum technologies; notably, quantum information processing and quantum metrology have been recognized to possibly benefit from non-Markovian noise sources \cite{Bellomo2007,Bylicka2014,Chin2012,Maniscalco2008}. 
 
\begin{figure}[t!]
	\centering
	\includegraphics[width=.47\textwidth]{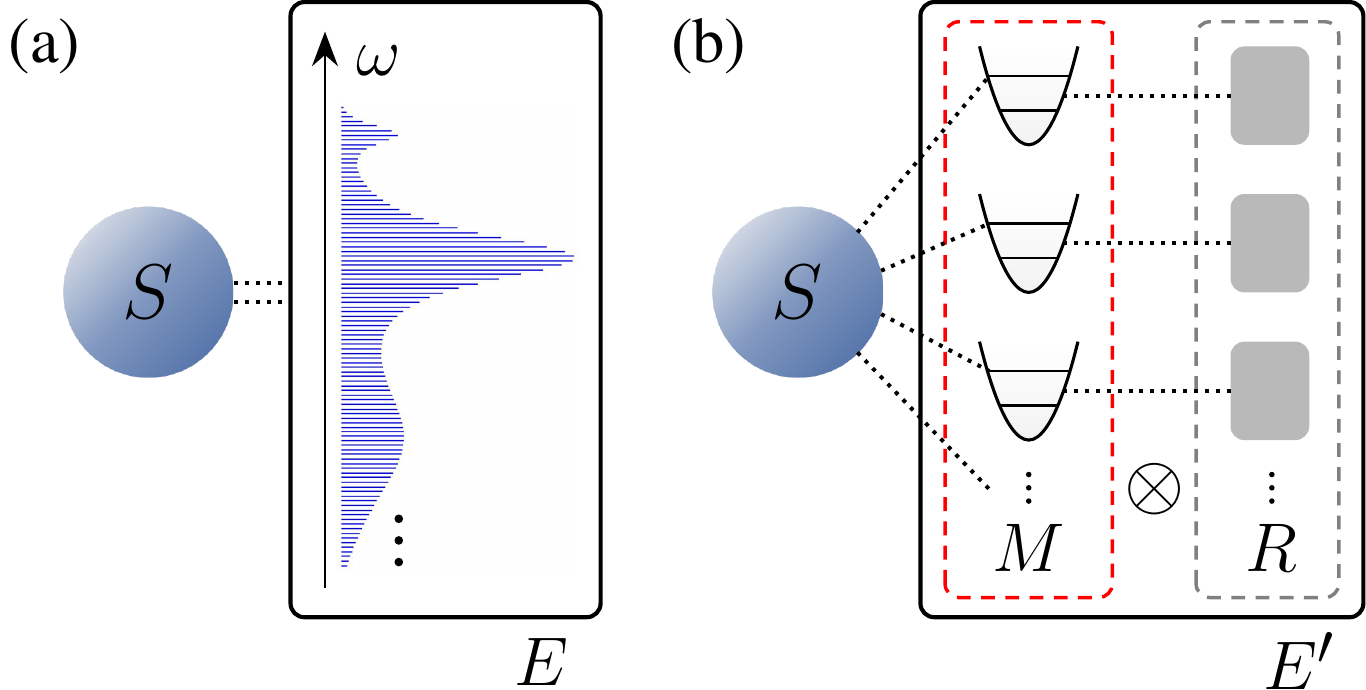}
	\caption{\label{fig:1} Schematic diagram showing the two considered models of the system-reservoir interaction. (a) An OQS $S$ interacting with a bosonic environment $E$. (b) The same system $S$ interacting with an auxiliary environment $E'$, consisting of a finite number of discrete modes $M$ coupled to local Markovian reservoirs $R$.}
\end{figure}

Within the validity of the Markov and weak-coupling (Born) approximations, it is well known that the quantum master equation describing the reduced dynamics can be generally expressed in Lindblad (GKSL) form \cite{Gorini1976,*Lindblad1976}. Master equations of this type have long been the focus of both theoretical and experimental research, not only because of their ability to describe essential features of dissipation and decoherence, but also due to the existence of efficient numerical methods for their solution \cite{Dalibard1992,Plenio1998,Gisin1992}. By contrast, while it is possible to derive a generalized form of master equation without the use of such approximations \cite{Nakajima1958,*Zwanzig1960}, the resulting non-Markovian equations of motion are often far too demanding to solve for an exact description to be feasible. For this reason a certain class of methods for treating complex open system problems have relied on the alternative idea of mapping the initial system onto a simpler, auxiliary configuration, consisting of the original open system coupled to a small number of auxiliary bosonic (fermionic) modes, which in turn are coupled to an external Markovian reservoir \cite{Imamoglu1994,Stenius1996,Chin2010,*Woods2014,*Tamascelli2019,ISmith2014,*ISmith2016,Strasberg2016,Roden2011,*Roden2012,Schonleber2015,Arrigoni2013,*Dorda2013,*Dorda2017,Pleasance2017,Garraway1997,Dalton2001,Dalton2003,*Garraway2006} (see Fig. \ref{fig:1}). In particular, the pseudomode method has effectively utilized such a mapping to describe the non-Markovian dynamics of a two-level system interacting with a bosonic environment \cite{Pleasance2017,Garraway1997}. In this approach the environment is replaced by a set of auxiliary discrete modes\textemdash the pseudomodes\textemdash which are identified through evaluating the poles of the spectral density function (i.e., the Fourier transform of the reservoir correlation function) when analytically continued to the lower-half complex frequency plane. By expanding the system over the pseudomodes, one can derive a Lindblad master equation describing the dynamics induced by the non-Markovian interaction between the pseudomodes and two-level system, in addition to the coupling of the pseudomodes to an external Markovian environment. Importantly, while this method is exact, its application is currently restricted to regimes where only one excitation is initially present in the system, as well as to interactions valid within the rotating wave approximation.

Beyond this approach, we note in Ref. \cite{Dalton2001} that a similar type of mapping has been employed in conjunction with the Fano diagonalization technique to extend the treatment of the pseudomode method to multiple excitation regimes. The method is distinct from \cite{Garraway1997} in that it instead relies on ``undressing" the environment into a set of auxiliary quasimodes, whose parameters\textemdash including the couplings to the system and overall configuration (i.e., site energies and intermode couplings)\textemdash are chosen so as to recover the spectral density function of the original environment. However, owing to the general difficulty of determining these parameters exactly, applications of the mapping so far have only focused on specific models where the system of interest is either coupled to a high-$Q$ cavity or photonic band-gap reservoir \cite{Dalton2003,*Garraway2006}. It is also worth noting that a number of related mappings have been put forward in the literature \cite{Chin2010,*Woods2014,*Tamascelli2019,ISmith2014,*ISmith2016,Strasberg2016}. The approaches outlined in Refs. \cite{Imamoglu1994,Stenius1996,Arrigoni2013,*Dorda2013,*Dorda2017}, for example, rely on replacing the physical environment by an \textit{ad hoc} collection of discrete modes whose parameters are fitted as those which most accurately represent the spectral density of the original model. Although these mappings can be applied quite generally, unlike \cite{Garraway1997,Dalton2001}, they have the disadvantage of not always being grounded in exact relations between the physical and auxiliary environments, requiring their accuracy to often be validated against exact numerical techniques. 

Recently, a proof of an exact mapping of a non-Markovian open system onto a Markovian one for a system interacting with a Gaussian (bosonic) environment was given in \cite{Tamascelli2018} (c.f. also Refs. \cite{Chen2019,Lambert2019}). There it was shown that the reduced dynamics of a non-Markovian system can be equivalently described in terms of an exact Lindblad-type master equation for a enlarged Markovian open system (system plus discrete modes), which in the context of the pseudomode mapping was used to generalize Ref. \cite{Garraway1997} beyond single-excitation regimes for cases where the reservoir spectral density is Lorentzian. In this paper we extend this treatment to instances where the exact dynamics of the enlarged system is described by a non-Lindblad form of master equation with a \textit{non-Hermitian} interaction Hamiltonian, as well as to physical environments modeled by a generalized spectral density function. To achieve this we explicitly generalize the proof given in Ref. \cite{Tamascelli2018} to account for an auxiliary environment which may admit a non-Hermitian coupling to the system. In particular, our approach relies on the connection between the poles of the spectral density function in the lower-half complex plane, and the properties of the discrete modes used represent the memory part of the environment. For certain spectral density functions we further show how the master equation may be brought into an appropriate Lindblad form by applying an effective change of basis to the discrete modes.

This paper is organized as follows. After outlining the physical model in Sec. \ref{sec:2}, we proceed Sec. \ref{sec:3} to introduce a mapping of the initial problem onto an auxiliary model and subsequently prove the reduced system dynamics to be indistinguishable between the two. In Sec. \ref{sec:4} we then derive an exact form of master equation for the enlarged system and present an initial application of this result. In Sec. \ref{sec:5} we address for certain cases how to convert between pathological (non-Lindblad) and Lindblad forms of the derived master equation. Finally, a summary and outlook is presented in Sec. \ref{sec:6}.

\section{Physical model}\label{sec:2}

We start by considering a generic microscopic model of an open quantum system (OQS) $S$ interacting with a bosonic environment $E$, as depicted in Fig. \ref{fig:1}(a). The total Hamiltonian of the model is written as
\begin{equation}\label{eq:H}
	H=H_S+H_E+H_I,
\end{equation}
where $H_S$ and $H_E$ are the Hamiltonians of the system and environment acting on the respective (Hilbert) subspaces $\mathcal{H}_S$ and $\mathcal{H}_E$, with $H_I$ an interaction term describing the effects of the system-environment coupling on $\mathcal{H}_S\otimes\mathcal{H}_E$. The Hamiltonians $H_E$ and $H_I$ are given by ($\hbar=1$)
\begin{align}
	H_{E} &= \sum_{\lambda}\omega_{\lambda}a^{\dagger}_{\lambda}a_{\lambda},\label{eq:H_E}\\
	H_I &= \sum_{j,\lambda}\left(g_{j\lambda}c^{\dagger}_ja_{\lambda}+\text{h.c.}\right).\label{eq:H_I}
\end{align}
Here $a_{\lambda}$ ($a^{\dagger}_{\lambda}$) is the bosonic annihilation (creation) operator for an excitation of frequency $\omega_{\lambda}$ satisfying the usual commutation relation $[a_{\lambda},a^{\dagger}_{\lambda'}]=\delta_{\lambda\lambda'}$, $c_j$ ($c^{\dagger}_j$) is a generic OQS operator associated to the $j$ transition of $S$ involved in the coupling, and $g_{j\lambda}$ denotes the coupling strength between the $\omega_{\lambda}$ mode of the field and the $j$ transition of the OQS. \\
\indent In what follows the system Hamiltonian $H_S$ is to be left unspecified and may in general have an explicit time dependence. On the other hand, the free evolution of the OQS (i.e., the evolution occurring in the absence of any driving or coupling between internal degrees of freedom) is described by the Hamiltonian
\begin{equation}\label{eq:H^0_S}
	H_{S,0}=\sum^{d_S}_{n=1}\epsilon_n\ket{e_n}\bra{e_n},
\end{equation}
with the set of discrete energy levels (eigenenergies) of the system denoted by $\{\ket{e_n}\}_S$ ($\epsilon_n$) and $d_S=\text{dim}\,\mathcal{H}_S$. The transition (jump) operators $c_j$, $c^{\dagger}_j$ appearing in Eq. (3) are formally defined as \cite{BreuerTOQS2002}
\begin{equation}
	c_j = \sum_{\epsilon_m-\epsilon_n=\omega_j}\Pi(\epsilon_n)O_j\Pi(\epsilon_m),
\end{equation}
where $O_j$ is a system observable and $\Pi(\epsilon_n) = \ket{e_n}\bra{e_n}$. Based on this definition, we have that the OQS transition operators satisfy the eigenoperator relations $[H_{S,0},c_j]=-\omega_jc_j$ ($[H_{S,0},c^{\dagger}_j]=\omega_jc^{\dagger}_j$). Hence $c_j$ ($c^{\dagger}_j$) lowers (raises) the internal energy of the OQS by an amount $\omega_j$. Moving to an interaction picture generated by the unitary transformation
\begin{equation}\label{eq:U_0}
	U_0(t)=\text{exp}[-i(H_{S,0}+H_E)t], 
\end{equation}
the interaction Hamiltonian $H_I$ becomes 
\begin{align}\label{eq:H_I_int}
	H_I(t)&=\sum_{j,\lambda}\left(g_{j\lambda}c^{\dagger}_ja_{\lambda}e^{-i(\omega_{\lambda}-\omega_j)t}+\text{h.c.}\right)\nonumber\\
		 &\equiv\sum_j\left[c^{\dagger}_j(t)\otimes B_j(t)+c_j(t)\otimes B^{\dagger}_j(t)\right],
\end{align}
with
\begin{equation}\label{eq:B_j}
	B_j(t) = \sum_{\lambda}g_{j\lambda}a_{\lambda}e^{-i\omega_{\lambda}t}
\end{equation}
defining the environmental noise operators. Notice in particular that the absence of terms oscillating at frequencies $\pm i(\omega_{\lambda}+\omega_j)$ in Eq. (\ref{eq:H_I_int}) implies the use of the rotating wave approximation (RWA). \\
\indent Following Ref. \cite{Tamascelli2018}, we are interested in examining the time-dependent behavior of the OQS in cases involving initially factorizing conditions $\rho_{SE}(0)=\rho_S(0)\otimes\rho_E(0)$, where for simplicity the environment is taken at $t=0$ to be in the vacuum state: 
\begin{equation}\label{eq:rho_E}
	\rho_E(0)=\ket{0}\bra{0}_E.
\end{equation}
Because $\rho_E(0)$ is then Gaussian and satisfies $\text{Tr}_E[B_j(t)\rho_E(0)]=0$, the OQS dynamics described by the reduced density operator 
\begin{equation}\label{eq:reduced_S}
	\rho_S(t)=\text{Tr}_E[\rho_{SE}(t)]
\end{equation}
will only depend on the second-order moments of the noise operators $B_j(t)$, $B^{\dagger}_j(t)$. For this model, these are explicitly written in terms of the two-time correlation functions ($t\geq s\geq0$)
\begin{align}\label{eq:zero_corr}
	\langle B^{\dagger}_j(t)B_k(s)\rangle_E &= 0,\nonumber\\
	\langle B_{j}(t)B_k(s)\rangle_E &= 0 = \langle B^{\dagger}_j(t)B^{\dagger}_k(s)\rangle_E,\quad\forall j,k,
\end{align}
with $\langle\cdot\rangle_E\equiv\text{Tr}_E[\,\cdot\,\rho_E(0)]$, and
\begin{align}\label{eq:2tcorrE}
	f_{jk}(t-s) &\equiv \langle B_j(t)B^{\dagger}_k(s)\rangle_E\nonumber\\
	           &=\sum_{\lambda}g_{j\lambda}g^*_{k\lambda}e^{-i\omega_{\lambda}(t-s)}.
\end{align}
The continuum limit of Eq. (\ref{eq:2tcorrE}) can now be taken by replacing the sum over the coupling constants $g_{j\lambda}$ with an integral weighted by the density of states $\rho_{\lambda}$ of the reservoir modes. Since the only quantities entering into the physical description are $\rho_{\lambda}$ and $|g_{j\lambda}|^2$, we may combine their joint frequency dependence into a single spectral density function $D(\omega_{\lambda})$,
\begin{equation}\label{eq:D}
	\rho_{\lambda}|g_{j\lambda}|^2 = \frac{\Omega^2_j}{2\pi}D(\omega_{\lambda}),
\end{equation}
with $D(\omega)$ normalized to
\begin{equation}\label{eq:norm}
	\int^{\infty}_{-\infty} d\omega D(\omega) = 2\pi,
\end{equation}
so that in turn, the quantities $\Omega_j$ act to measure the coupling strength of the $j$ transition of the OQS to the full set of environment modes via the expression
\begin{equation}\label{eq:Omega_j}
	\Omega^2_j = \int d\omega_{\lambda}\,\rho_{\lambda}|g_{j\lambda}|^2.
\end{equation} 
\indent For the remainder of this paper our focus will be on describing the non-Markovian dynamics of OQSs coupled to various types of structured reservoir. Hence, for this purpose we shall model the system-reservoir interaction (\ref{eq:H_I_int}) using a generalized form of spectral density function which may vary strongly over $\omega$ with respect to the frequency scales of the OQS. In particular, it will only be assumed that $D(\omega)$ is a meromorphic function when analytically continued to the lower-half complex $\omega$ plane and that $D(\omega)$ tends to zero at least as fast as $\sim O(1/|\omega|^2)$ for $|\omega|\rightarrow\infty$. Under these assumptions, and with all other nonanalytic features of the spectral density function removed (e.g., branch cuts), the two-time correlation function (\ref{eq:2tcorrE}) may then be evaluated solely in terms of the poles and residues of $D(\omega)$ via contour integration methods. In this way we proceed to write Eq. (\ref{eq:2tcorrE}) in terms of the integral
\begin{equation}\label{eq:cont_int}
	f_{jk}(t-s) = -\frac{\Omega_j\Omega_k}{2\pi}\oint_{C}d\omega D(\omega)e^{-i\omega(t-s)}, 
\end{equation} 
where $C$ is a contour defined along the full real line and closed by a semicircular arc in the lower-half complex plane. We note that the construction of (\ref{eq:cont_int}) relies on the use of the RWA, which in the limit $\omega_j\rightarrow\infty$ (or, more loosely, for $\omega_j\gg \Omega_j,\lambda_l,|\Delta_{jl}|$) formally allows an extension of the environment definition to include modes of negative frequency $\omega<0$. Furthermore, the poles of $D(\omega)$ in the lower-half complex $\omega$ plane are located at positions $z_{1},z_{2},\dotso z_l,\dotso$ with their corresponding residues denoted by $r_{1},r_{2},\dotso r_l,\dotso$, while each $z_l$ has real and imaginary parts
\begin{equation}\label{eq:pole}
	z_l=\xi_l-i\lambda_l.
\end{equation}
We can now apply the residue theorem to Eq. (\ref{eq:cont_int}) to obtain ($\tau\equiv t-s$)
\begin{equation}\label{eq:2tcorrE_res}
	f_{jk}(\tau)=-i\Omega_j\Omega_k\sum_lr_le^{-iz_l\tau},\quad\tau\geq0,
\end{equation}
where in the following it will also prove useful to define the coupling constants
\begin{equation}\label{eq:g_jl}
	g'_{jl}\equiv\Omega_j\sqrt{-ir_l},
\end{equation}
which are in general complex quantities. For simplicity we shall first restrict ourselves to real couplings $g'_{jl}$. Besides the assumptions already made on the spectral density (\ref{eq:D}), this imposes no extra limitations on the model given that $f_{jk}(0)=-i\Omega_j\Omega_k\sum_lr_l$ must always evaluate to the real quantity $\Omega_j\Omega_k$ [c.f. Eqs. (\ref{eq:2tcorrE})-(\ref{eq:Omega_j})], and hence $(-ir_l)$ has no net imaginary part, i.e.,
\begin{equation}\label{eq:norm_g}
	\sum_l(-ir_l)=1. 
\end{equation}
The most general case involving complex couplings $g'_{jl}$ will be considered later on in Sec. \ref{sec:5}.

\section{Auxiliary model}\label{sec:3}

In this section we proceed to introduce the auxiliary model that will allow us to represent the reduced evolution of Eq. (\ref{eq:reduced_S}) within an enlarged open system whose dynamics is Markovian. To this end, let us first consider a mapping of Eqs. (\ref{eq:H_E}) and (\ref{eq:H_I}) in which the original Hamiltonian (\ref{eq:H}) is replaced by
\begin{align}\label{eq:H'}
	H&=H_S+H_E+H_I\nonumber\\
	   &\rightarrow H'=H_S+H_{E'}+H'_I.
\end{align}
The mapping modifies the environment $E$ so that the open system $S$ is now coupled to a set of auxiliary discrete modes $M$, which in turn are each coupled to an independent reservoir with vanishing correlation time [c.f. Fig. \ref{fig:1}(b)]. Thus the Hamiltonian of the new environment configuration $E'$ reads 
\begin{align}
	H_{E'} & = H_M + H_R + H_{MR},\nonumber\\
	H_R &= \sum_l\int^{\infty}_{-\infty} d\omega\,\omega a^{\dagger}_{Rl}(\omega)a_{Rl}(\omega),\label{eq:H_R}\\
	H_{MR} &= \sum_l\sqrt{\frac{\lambda_l}{\pi}}\int^{\infty}_{-\infty} d\omega\Big(b^{\dagger}_la_{Rl}(\omega)+\text{h.c.}\Big),\label{eq:H_MR}
\end{align} 
where $b_l$ ($b^{\dagger}_l$) is the annihilation (creation) operator for a discrete bosonic mode $M_l$, $a_{Rl}(\omega)$ ($a^{\dagger}_{Rl}(\omega)$) is the annihilation (creation) operator for an excitation of frequency $\omega$ in the reservoir $R_l$, and $[b_l,b^{\dagger}_{l'}]=\delta_{ll'}$, $[a_{Rl}(\omega),a^{\dagger}_{Rl}(\omega')]=\delta(\omega-\omega')$. The free Hamiltonian of the full set of discrete modes $M$ is written as
\begin{equation}\label{eq:H_M}
	H_M = \sum_l\xi_lb^{\dagger}_lb_l,
\end{equation}
while their coupling to the system $S$ is described by
\begin{equation}\label{eq:H_I'}
	H'_I = \sum_{j,l}g'_{jl}\left(c^{\dagger}_j\otimes b_l+\text{h.c.}\right).
\end{equation} 
Moreover, we stress that the parameters of the Hamiltonians (\ref{eq:H_MR}) and (\ref{eq:H_M}) have been chosen in such a way that the discrete modes have the same one-to-one association with the poles of $D(\omega)$ as the pseudomodes introduced in Ref. \cite{Garraway1997}. \\
\indent Our next step is to transform $H'_I$ to the same interaction picture as Eq. (\ref{eq:H_I_int}) by means of the unitary operator 
\begin{equation}
	U'_0(t) = \text{exp}[-i(H_{S,0}+H_{E'})t], 
\end{equation}
which following the technique of Eq. (\ref{eq:H'}), has simply been obtained by replacing the free Hamiltonian $H_E$ in Eq. (\ref{eq:U_0}) with $H_{E'}$. The interaction Hamiltonian in this frame of reference reads
\begin{align}\label{eq:H_I'_int}
	H'_I(t) &= \sum_{j,l}g'_{jl}\left(c^{\dagger}_j\otimes b_l(t)e^{i\Delta_{jl}t}+c_j\otimes b^{\dagger}_l(t)e^{-i\Delta_{jl}t}\right)\nonumber\\
		     &\equiv \sum_j\left[c^{\dagger}_j(t)\otimes B'_j(t) + c_j(t)\otimes B'^{\dagger}_j(t)\right],
\end{align}
with detunings $\Delta_{jl}=\omega_j-\xi_l$ from the OQS transition frequency $\omega_j$, $b_j(t)=U^{\dagger}_{MR}(t)b_j(0)U_{MR}(t)$, and 
\begin{equation}\label{eq:U_MR}
	U_{MR}(t,0) = \mathcal{T}\text{exp}\left[-i\int^t_0ds\,H_{MR}(s)\right].
\end{equation}
Here $U_{MR}(t,0)\equiv U_{MR}(t)$ is a unitary operator describing the time evolution of the free environment oscillators (i.e., the degrees of freedom of $M+R$ with no coupling to the OQS), $\mathcal{T}$ is the chronological time-ordering operator \cite{BreuerTOQS2002}, and
\begin{equation}\label{eq:H_MR(t)}
	H_{MR}(t) = \sum_l\sqrt{\frac{\lambda_l}{\pi}}\int d\omega\left(b^{\dagger}_la_{Rl}(\omega)e^{-i(\omega-\xi_l)t}+\text{h.c.}\right).
\end{equation}
Note that we have also defined the general form of noise operator $B'_j(t)$, which in the auxiliary model is the counterpart to that given in Eq. (\ref{eq:B_j}): 
\begin{equation}
	B'_j(t) = \Omega_j\sum_l\sqrt{-ir_l}\,b_l(t)e^{-i\xi_lt}.
\end{equation}
Now, fixing the environment $M+R$ to have the same initial conditions as $E$ above, namely, by choosing an initially factorized state $\rho_{SMR}(0)=\rho_S(0)\otimes\rho_{MR}(0)$, and for $\rho_{MR}(0)$ to be given as 
\begin{align}\label{eq:rho_MR}
	\rho_{MR}(0) &= \ket{0}\bra{0}_M\otimes\ket{0}\bra{0}_R,
\end{align}
with $\ket{0}_{M/R}\equiv\bigotimes_l\ket{0}_{M_l/R_l}$, the expectation values of the environmental noises satisfy $\text{Tr}_{MR}[B'_j(t)\rho_{MR}(0)]=0$. This is based on $\rho_{MR}(0)$ being a stationary state of Eq. (\ref{eq:U_MR}), i.e.,
\begin{equation}\label{eq:stationary}
	[H_{MR}(t),\rho_{MR}(0)]=0.
\end{equation}
Hence, in an analogous way to Eq. (\ref{eq:reduced_S}), the reduced dynamics of
\begin{equation}\label{eq:reduced_S_aux}
	\rho'_S(t) = \text{Tr}_{MR}[\rho_{SMR}(t)]
\end{equation}
will only depend on the second-order moments of the environmental noise operators $B'_j(t)$, $B'^{\dagger}_j(t)$. The only nonzero contribution written in terms of these moments is ($t\geq s$)
\begin{align}\label{eq:2tcorrE'}
	&\langle B'_j(t)B'^{\,\dagger}_k(s)\rangle_{E'} = -i\Omega_j\Omega_k\sum_{i,l}\sqrt{r_ir_l}\,e^{-i(\xi_lt-\xi_is)}\nonumber\\
	&\,\,\times\text{Tr}_{MR}\left[U^{\dagger}_{MR}(t,s)b_lU_{MR}(t,s)b^{\dagger}_i\rho_{MR}(0)\right]\nonumber\\
	&\equiv-i\Omega_j\Omega_k\sum_{i,l}\sqrt{r_ir_l}\,e^{-i(\xi_lt-\xi_is)}\langle b_l(t,s)b^{\dagger}_i(0)\rangle_{E'},
\end{align}
since for our initial choice of state satisfying Eq. (\ref{eq:stationary}) and $b_l\ket{0}_M=0$ $\forall l$, all other correlation functions either quadratic in $b_l$ and $b^{\dagger}_l$ or proportional to $\langle b^{\dagger}_l(t,s)b_i(0)\rangle_{E'}$ can be shown to vanish in line Eq. (\ref{eq:zero_corr}).

\subsection{Comparison between the OQS dynamics of the physical and auxiliary models}

With the relevant details in place, we now look to prove an equivalence between the OQS dynamics generated by the two forms of interaction in Eqs. (\ref{eq:H_I}) and (\ref{eq:H_I'}). Based on the discussion so far, our proof exploits the fact that given the initial choice of vacuum states (\ref{eq:rho_E}) and (\ref{eq:rho_MR}), the reduced system dynamics of the two models will be identical as long as the two-time correlation functions of the physical and auxiliary environments share the same time dependence \cite{Tamascelli2018,Stenius1996}. Thus, using Eq. (\ref{eq:2tcorrE'}) as our starting point, it remains for us to then solve the Heisenberg equations of motion for the operators $b_l(t,s)$ and use this result to obtain a time-dependent expression for $\langle b_l(t,s)b^{\dagger}_i(0)\rangle_{E'}$. \\
\indent In a Heisenberg picture generated via Eq. (\ref{eq:U_MR})\textemdash that is, where $b_l(t,s)=U^{\dagger}_{MR}(t,s)b_lU_{MR}(t,s)$\textemdash we solve the corresponding operator equations of motion for $b_l(t,s)$ in Appendix \ref{appenA} to obtain
\begin{equation}\label{eq:b_l(t)}
	b_l(t,s) = e^{-\lambda_l(t-s)}b_l(0)-i\sqrt{2\lambda_l}\int^t_sdt_1e^{-\lambda_l(t-t_1)}a^{\text{in}}_{Rl}(t_1).
\end{equation}
Note that here we have introduced the noise operators
\begin{equation}\label{eq:a_in}
	a^{\text{in}}_{Rl}(t)= \frac{1}{\sqrt{2\pi}}\int^{\infty}_{-\infty} d\omega\,a_{Rl}(\omega)e^{-i(\omega-\xi_l)t},
\end{equation}
which adopt the same definition as the so-called ``input" fields introduced in the Gardiner-Collet description of (Markovian) quantum white noise \cite{Gardiner2005,Gardiner1985}. Accordingly, since the reservoirs $R_l$ are each initially taken to be in the vacuum states $\rho_{R_l}(0)=\ket{0}\bra{0}_{R_l}$, the expectation values of the noise operators satisfy $\text{Tr}_{R_l}\left[a^{\text{in}}_{Rl}(t)\rho_{R_l}(0)\right]=0$. This allows us to directly substitute Eq. (\ref{eq:b_l(t)}) into (\ref{eq:2tcorrE'}) to obtain
\begin{align}\label{eq:2tcorrE_equiv}
	\langle B'_j(t)B'^{\,\dagger}_k(s)\rangle_{E'}&=-i\Omega_j\Omega_k\sum_lr_le^{-iz_l(t-s)}\nonumber\\
								 	 &=f_{jk}(\tau),\quad \forall j,k,\,\tau\geq0,
\end{align}
which proves the full equivalency of the two-time correlation functions of $E$ and $M+R$, given that all other correlation functions equally match due to having a trivial time dependence. Because the environmental noise operators $B_j(t)$ and $B'_j(t)$ have then been shown to be identical correlation-wise, we may therefore conclude that the dynamics of the OQS are \textit{indistinguishable} between the two models. In other words, the non-Markovian response of the system $S$ is invariant under replacing the physical environment $E$ by a finite number of discrete modes $M$ coupled to independent Markovian reservoirs $R$; from Eqs. (\ref{eq:reduced_S}) and (\ref{eq:reduced_S_aux}), it subsequently follows that
\begin{equation}\label{eq:equiv}
	\rho'_S(t)=\rho_S(t).
\end{equation}
Finally, we restate our main assumptions of $D(\omega)$ being a meromorphic function in the lower-half complex $\omega$ plane and of the coupling constants $g'_{jl}$ being real.

\section{Exact solution to the problem}\label{sec:4}

Dealing with the auxiliary model in place of Eq. (\ref{eq:H}) now enables us to reproduce the exact OQS dynamics without making any form of approximation involving weak coupling or separation of timescales between the system and environment. To show this explicitly, we will proceed to derive the quantum Langevin equation for the enlarged open system comprising the original system $S$ and the discrete modes $M$. For convenience, we choose to work in an interaction picture with respect to the free Hamiltonian $H_0=H_{S,0}+H_M+H_R$. In this frame of reference, the time evolution for an arbitrary operator $A$ of the enlarged system is defined $A(t) = U^{\dagger}(t)A(0)U(t)$, where the unitary operator $U(t)$ satisfies the Schr\"{o}dinger equation $\frac{d}{dt}U(t)=-iH(t)U(t)$, with
\begin{equation}\label{eq:H(t)}
	H(t) = H_S(t) + H_I(t) + H_{MR}(t),
\end{equation}
and 
\begin{equation}\label{eq:H_sys_dm}
	H_I(t) = \sum_{j,l}g'_{jl}\left(c^{\dagger}_j\otimes b_l\,e^{i\Delta_{jl}t}+\text{h.c.}\right).
\end{equation}
Notice that since the two configurations of environment $E$ and $M+R$ are interchangeable at the level of the OQS, here we have dropped the dash label used to distinguish the interaction Hamiltonian $H'_I(t)$ against that of the physical model (\ref{eq:H_I_int}). Furthermore, the system Hamiltonian reads $H_S(t) = \text{exp}(iH_0t)(H_S-H_{S,0})\text{exp}(-iH_0t)$, where $H_{MR}(t)$ again is given by Eq. (\ref{eq:H_MR}). \\
\indent The Heisenberg equation of motion for an arbitrary operator $A(t)$ is written as 
\begin{align}\label{eq:HE_A}
	\frac{d}{dt}A(t) &= -i[A(t),H_S(t)+H_I(t)]\nonumber\\
	&-i\sum_l\sqrt{\frac{\lambda_l}{\pi}}\int d\omega\Big(a^{\dagger}_{Rl}(\omega,t)e^{i(\omega-\xi_l)t}[A(t),b_l(t)]\nonumber\\
	&\,+[A(t),b^{\dagger}_l(t)]a_{Rl}(\omega,t)e^{-i(\omega-\xi_l)t}\Big),
\end{align}
so that by formally eliminating the reservoir variables $a_{Rl}(\omega,t)=U^{\dagger}(t)a_{Rl}(\omega)U(t)$ from Eq. (\ref{eq:HE_A}) one obtains the quantum Langevin equation
\begin{widetext}
\begin{align}\label{eq:QLE_A}
	\frac{d}{dt}A(t) = &-i[A(t),H_S(t)+H_I(t)]+\sum_l\lambda_l\left\{b^{\dagger}_l(t)[A(t),b_l(t)]-[A(t),b^{\dagger}_l(t)]b_l(t)\right\}\nonumber\\
			          &-i\sum_l\sqrt{2\lambda_l}\left\{[A(t),b^{\dagger}_l(t)]a^{\text{in}}_{Rl}(t)+a^{\text{in}\,\dagger}_{Rl}(t)[A(t),b_l(t)]\right\}.
\end{align}
\end{widetext}
At this stage we may derive the master equation for the reduced density matrix $\rho(t)\equiv\text{Tr}_R[\rho_{SMR}(t)]$ by taking the expectation value of both sides of Eq. (\ref{eq:QLE_A}) with respect to an initially factorized density matrix $\rho_{SMR}(0)=\rho(0)\otimes\rho_R(0)$ ($\rho_R(0)=\ket{0}\bra{0}_R)$. Since the noise terms proportional to $a^{\text{in}}_{Rl}(t)$ and $a^{\text{in}\,\dagger}_{Rl}(t)$ do not contribute from 
 \begin{equation}\label{eq:b_in_zero}
	a^{\text{in}}_{Rl}(t)\rho_R(0)=0=\rho_R(0)a^{\text{in}\,\dagger}_{Rl}(t),
\end{equation}
the remaining commutators are easily expanded to obtain
\begin{align}\label{eq:aveHE_2}
 	&\frac{d}{dt}\left\langle A(t)\right\rangle = -i\left\langle[A(t),H_S(t)+H_I(t)]\right\rangle_{SMR}\nonumber\\
	&+\sum_l\lambda_l\left\langle b^{\dagger}_l(t)[A(t),b_l(t)] - [A(t),b^{\dagger}_l(t)]b_l(t)\right\rangle_{SMR}.
\end{align}
The resulting equation now is expressed solely in terms of operators pertaining to the enlarged OQS $S+M$. Hence one may use the cyclic trace property
\begin{equation}\label{eq:exp}
	\left\langle A(t)\right\rangle=\text{Tr}_{SM}\big[\text{Tr}_R[A(t)\rho(0)\otimes\ket{0}\bra{0}_R]\big]=\text{Tr}_{SM}\left[A\rho(t)\right]
\end{equation}
to move $A$ to the leftmost side of each term, e.g., $\text{Tr}_{SM}[Ab_l\rho(t)b^{\dagger}_l]$, thereby allowing us to read off each of the terms in the master equation for $\rho(t)$. Through doing so, we finally arrive at
\begin{equation}\label{eq:ME}
	\frac{d}{dt}\rho(t)=-i\left[H_S(t)+H_I(t),\rho(t)\right]+\mathcal{D}[\rho(t)],
\end{equation}
where the superoperators
\begin{equation}
	 \mathcal{D}[\rho]=2\sum_l\lambda_l\left(b_l\rho b^{\dagger}_l-\frac{1}{2}\big\{b^{\dagger}_lb_l,\rho\big\}\right)
\end{equation}
describe the local dissipation of each discrete mode occurring at rate $\lambda_l$. \\
\indent Equation (\ref{eq:ME}) establishes one of the main results of this paper. The master equation is of a standard Lindblad form \cite{Gorini1976,*Lindblad1976} and represents the general evolution of a reduced system density matrix $\rho_S(t)=\text{Tr}_M[\rho(t)]$ as the projection of a larger quantum Markov process, where all memory effects contained in the non-Markovian dynamics are incorporated into the coupling between the OQS and discrete modes. Importantly, this general property of Eq. (\ref{eq:ME}) also allows an unravelling of the master equation into Markovian pure state trajectories such that a numerically efficient simulation to the problem may be readily obtained using the quantum jump (Monte Carlo wave function) method \cite{Dalibard1992,*Plenio1998} or other related approaches \cite{Gisin1992}. In this respect our result may be considered as a non-Markovian generalization of quantum jump method. In fact, a number of recent studies have shown similar extensions of this approach to be possible by considering the evolution a non-Markovian OQS embedded within a larger Markov process \cite{Diosi2012,*Budini2013,*Breuer2004,*Mazzola2009}. 

\subsection{Application of the result}\label{sec:6A}

For the sake of concreteness let us briefly examine an application of our result. As a convenient example we consider a two-level system (TLS) $S$ with excited (ground) state $\ket{e}$ ($\ket{g}$) interacting with a bosonic environment $E$ at zero temperature. Within the RWA, the interaction Hamiltonian reads $H_I(t)=\sigma_+B(t)+\sigma_-B^{\dagger}(t)$, where $\sigma_+=\sigma^{\dagger}_-=\ket{e}\bra{g}$ denotes the Pauli matrices such that $\sigma_+\ket{g}$ is an eigenstate of the free Hamiltonian $H_{S,0}=\omega_0\sigma_+\sigma_-$ with eigenvalue $\omega_0$. The only nonzero correlation function of the environment is taken to be of the form of Eq. (\ref{eq:2tcorrE_res}),
\begin{align}\label{eq:2tcorrE_ex}
	 f(\tau) &= \text{Tr}_E\Big[B(\tau)B^{\dagger}(0)\rho_E(0)\Big]\nonumber\\
	           &= -i\Omega^2_0\sum^{N}_{l=1}r_le^{-iz_l\tau},\quad\tau\geq 0,
\end{align}
which as before is associated to real coupling constants $g'_l=\Omega_0\sqrt{-ir_l}$. \\
\indent To now find the exact master equation generated from the TLS-discrete-mode interaction, we can apply Eqs. (\ref{eq:H_sys_dm}) and (\ref{eq:ME}) to get ($\Delta_l=\omega_0-\xi_l$)
\begin{align}\label{eq:PMME}
	&\frac{d}{dt}\rho(t)=-i\left[H_S(t),\rho(t)\right]\nonumber\\
	&\,-i\sum^{N}_{l=1}g'_l\left[e^{i\Delta_lt}\sigma_+b_l+e^{-i\Delta_lt}b^{\dagger}_l\sigma_-,\rho(t)\right]+\mathcal{D}[\rho(t)].
\end{align}
The result agrees with that recently obtained by Tamascelli \textit{et al.} in Ref. \cite{Tamascelli2018} within the RWA. Moreover, when the total system contains at most a single excitation, i.e., $H_S(t)=0$, the above reduces to the same form of master equation derived via the pseudomode method \cite{Garraway1997}, where for this case each discrete mode adopts the role of a pseudomode. Therefore, not only have we recovered the result of Ref. \cite{Garraway1997} within the restrictions outlined above, but we have also generalized it via Eq. (\ref{eq:ME}) to a much greater variety of open system models, including those applying to the description of multiple excitation dynamics. \\ 
\indent To further illustrate the connection of our approach with the pseudomode theory, we note that one may obtain the equivalent spectral density to Eq. (\ref{eq:2tcorrE_res}) by inverting the general expression for the correlation function in Eq. (\ref{eq:cont_int}). Since $D(\omega)$ is by definition a real function, $f_{jk}(\tau)$ must be Hermitian in time, 
\begin{equation}
	f_{jk}(\tau)=f^*_{jk}(-\tau),\quad \tau\geq0,
\end{equation}
and so 
\begin{equation}\label{eq:D_Lorz}
	D(\omega) = 2\sum^N_{l=1}\frac{\text{Re}[r_l](\omega-\xi_l) + \lambda_l\text{Im}[r_l]}{(\omega - \xi_l)^2 + \lambda^2_l}. 
\end{equation}
In the case of real couplings $g'_{jl}$, i.e., $\text{Re}[r_l]=0$ and $\text{Im}[r_l]>0$, the spectral density function is found to reduce to a linear combination of Lorentzians, where each of the poles in the lower-half complex plane of $D(\omega)$ is connected to one of the $N$ discrete modes in the auxiliary model. Based on this association, we then find the discrete modes to be connected to precisely the same feature of the spectral density as the pseudomodes. On the other hand, it is also worth pointing out that the current restriction to real couplings $g'_{jl}$ limits the range of applicability of our theory to spectral density functions which can solely be written as a sum of positively weighted Lorentzians. This is illustrated by the fact that in the case of complex $g'_{jl}$, the form of spectral density obtained by inverting Eq. (\ref{eq:cont_int}) changes $D(\omega)$ from a sum of Lorentzians with positive weights to a more general form of function which is possibly either (i) non-Lorentzian or (ii) Lorentzian but with negative weights $\text{Im}[r_l]<0$, see (\ref{eq:D_Lorz}). Thus, if we are extend our generalization of the pseudomode method even further, it is clear we must look beyond the assumption of real couplings $g'_{jl}$.

\section{Non-Hermitian interaction Hamiltonian}\label{sec:5}

Remarkably, as is shown in Appendix \ref{appenB}, the considered proof does not actually rely on the assumption of $g'_{jl}$ being real\textemdash rather, the \textit{only} conditions necessary to guarantee the equivalence of $\rho_S(t)$ and $\rho'_{S}(t)$ is for the correlation functions of the physical and auxiliary environments to satisfy Eq. (\ref{eq:2tcorrE_equiv}). Therefore with the identity (\ref{eq:equiv}) still valid, one may proceed to write Eq. (\ref{eq:g_jl}) in terms of its real and imaginary parts,
\begin{equation}
	g'_{jl}\equiv g'^{(r)}_{jl} + ig'^{(i)}_{jl},
\end{equation}
so that the interaction Hamiltonian will comprise both a \textit{Hermitian} and \textit{anti-Hermitian} part:
\begin{equation}\label{eq:H_I'_nh}
	H_I(t) = H^{(r)}_I(t) + H^{(i)}_I(t).
\end{equation}
By now following the steps in the previous section we may again derive the master equation for the enlarged system. However, due to non-Hermitian nature of $H_I(t)$, we find the resulting master equation to no longer be of Lindblad form and in turn for the evolution of $\rho(t)$ to in general be nonpositive. Interestingly, we are then for this case able to generate a pathological form of master equation that, while is still capable of describing the correct OQS behavior, lacks a suitable physical interpretation for the dynamics of the enlarged system $S+M$. In particular, this may present issues for solving Eq. (\ref{eq:ME}) using quantum jump methods, given that the unravelling of $\rho(t)$ into quantum trajectories from the current form of master equation would admit the possibility of jumps occurring between states of the discrete-mode system with probabilities exceeding unity, among other unphysical defects. Below we shall focus on how the master equation may be brought into Lindblad form by applying a suitable transformation to the discrete modes.

\subsection{Lindblad construction of the master equation}\label{sec:5A}

The same issues connected with replacing an environment by a discrete set of bosonic modes which have complex coupling constants to the OQS have also previously been encountered with the pseudomode method \cite{Garraway1997}. In that context, i.e., with a RWA form of interaction and single excitations, a procedure was established to convert between pathological and Lindblad forms of the master equation by applying an effective change of basis to the pseudo (discrete) mode operators $b_l$ ($b^{\dagger}_l$). Hence, in the spirit of Ref. \cite{Garraway1997}, we enact a similar procedure below by first rewriting Eq. (\ref{eq:aveHE_2}) in the Schr\"{o}dinger picture:
\begin{align}
	\dot{\rho}(t) &= -i\left[H_{\text{eff}},\rho(t)\right] + 2\sum_l\lambda_lb_l\rho(t)b^{\dagger}_{l},\label{eq:ME_eff_1}\\
	H_{\text{eff}} &= H_S + \sum_lz_lb^{\dagger}_lb_l + \sum_{j,l}g'_{jl}\left(c^{\dagger}_j\otimes b_l+\text{h.c.}\right).\label{eq:H_eff}
\end{align}
Next we introduce a new set of discrete-mode operators $\tilde{b}_m$ ($\tilde{b}^{\dagger}_m$) via
\begin{equation}\label{eq:b_new}
	b_l = \sum_mU_{ml}\tilde{b}_m,
\end{equation}
with $U_{ml}$ an orthogonal (complex) matrix and $[\tilde{b}_m,\tilde{b}^{\dagger}_{m'}]=\delta_{mm'}$. Inserting the above decomposition into the interaction term of Eq. (\ref{eq:H_eff}) now leads to  
\begin{equation}
	\sum_{j,l}g'_{jl}\left(c^{\dagger}_j\otimes b_l+\text{h.c.}\right) = \sum_{j,m}\tilde{g}'_{jm}\left(c^{\dagger}_j\otimes\tilde{b}_m+\text{h.c.}\right),
\end{equation}
so that from $\tilde{g}'_{jm}=\sum_lU_{ml}g'_{jl}$, we can in principle fix the elements $U_{ml}$ of an otherwise arbitrary matrix by requiring the new couplings $\tilde{g}'_{jm}$ to be real. We also note with this definition that the couplings are constrained to satisfy the same normalization as Eq. (\ref{eq:norm_g}), i.e., $\sum_m(\tilde{g}'_{jm})^2=\Omega^2_j$. For the remaining terms, the effect of Eq. (\ref{eq:b_new}) will follow that of a similarity transformation applied to the diagonal matrices $z_l$ and $\lambda_l$, where
\begin{align}\label{eq:sim_trans_1}
	\sum_lz_lb^{\dagger}_lb_l &= \sum_l\sum_{m,m'}\left(U_{ml}z_lU^T_{lm'}\right)\tilde{b}^{\dagger}_m\tilde{b}_{m'}\nonumber\\
						&= \sum_{m,m'}\tilde{z}_{mm'}\tilde{b}^{\dagger}_m\tilde{b}_{m'},
\end{align}
and 
\begin{align}\label{eq:sim_trans_2}
	\sum_l\lambda_lb_l\rho(t)b^{\dagger}_l &= \sum_{m,m'}\left(U_{ml}\lambda_lU^T_{lm'}\right)\tilde{b}_{m'}\rho(t)\tilde{b}^{\dagger}_m\nonumber\\
								 &= \sum_{m,m'}\Gamma_{mm'}\tilde{b}_{m'}\rho(t)\tilde{b}^{\dagger}_m.
\end{align}
The non-Hermitian Hamiltonian $H_{\text{eff}}$ can subsequently be written in the form 
\begin{equation}
	H_{\text{eff}} = H_{SM} - i\sum_{m,m'}\Gamma_{mm'}\tilde{b}^{\dagger}_m\tilde{b}_{m'}
\end{equation}
with $\Gamma_{mm'}=-\text{Im}[\tilde{z}_{mm'}]$, and 
\begin{equation}
	H_{SM}= H_S + \sum_{m,m'}\text{Re}[\tilde{z}_{mm'}]\tilde{b}^{\dagger}_m\tilde{b}_{m'}+ \sum_{j,m}\tilde{g}'_{jm}\left(c^{\dagger}_j\tilde{b}_m+\text{h.c.}\right).
\end{equation}
Now, since $\text{Re}[\tilde{z}_{mm'}]$ is symmetric by definition, the master equation resulting from Eqs. (\ref{eq:sim_trans_1}) and (\ref{eq:sim_trans_2}) will only be in Lindblad form if we impose the additional constraint that the matrix $\Gamma_{mm'}$ is positive semidefinite \cite{Gorini1976,*Lindblad1976}. Under this restriction one can then obtain the following (Lindblad) master equation: 
\begin{align}\label{eq:ME_eff}
	&\frac{d}{dt}\rho(t) = -i\big[H_{SM},\rho(t)\big]\nonumber\\
	&\,\,+2\sum_{m,m'}\Gamma_{mm'}\left(\tilde{b}_{m'}\rho(t)\tilde{b}^{\dagger}_{m}-\frac{1}{2}\big\{\tilde{b}^{\dagger}_{m}\tilde{b}_{m'},\rho(t)\big\}\right).
\end{align}
Thus, while the OQS dynamics remains unaffected by the transformation, we find that the removal of pathological terms from Eq. (\ref{eq:ME}) generally introduces a nonzero coupling between the discrete modes. \\  
\indent Starting from Eq. (\ref{eq:ME_eff_1}), the task of converting between a non-Lindblad master equation and a Lindblad one now amounts to finding a transformation matrix $U_{ml}$ with the simultaneous requirements for $\tilde{g}'_{jm}$ to be real and for the dissipation (Kossakowski) matrix $\Gamma_{mm'}$ to be positive semi-definite. Unfortunately, for the general case involving $N$ discrete modes, this constitutes a highly nontrivial problem that has no guarantee of a unique solution. However, for the cases in which the OQS couples to only two discrete modes ($N=2$), it is in fact possible to obtain analytical expressions for $\tilde{z}_{mm'}$ and $\tilde{g}'_{jm}$ following directly the techniques of the pseudomode method \cite{Garraway1997}. Indeed, as determined from Sec. VB of that paper, one may here parametrize the matrix $\tilde{z}_{mm'}$ as ($m,m'=1,2$)
\begin{equation}\label{eq:z_km}
	\tilde{z}_{mm'} = (\tilde{\xi}_m-i\Gamma_m)\delta_{mm'}+V_{12}\left(\delta_{mm'-1}+\delta_{mm'+1}\right),
\end{equation}
where the analytical expressions for the new discrete-mode decay rates $\Gamma_m$, coupling constants $V_{12}=V^*_{12}$, and transition frequencies $\tilde{\xi}_m$ are given in Appendix \ref{appenC}.

\subsection{Further application of the result}

To provide an application of the theory outlined above, let us finally return to the OQS model introduced in Sec. \ref{sec:6A}\textemdash namely, of a TLS $S$ interacting with a zero-temperature bosonic field $E$\textemdash to consider the case in which the spectral density is instead given by the difference of two Lorentzian functions:
\begin{equation}\label{eq:PBG}
	D(\omega) = 2\left[W_1\frac{\lambda_1}{(\omega-\xi)^2+\lambda^2_1}-W_2\frac{\lambda_2}{(\omega-\xi)^2+\lambda^2_2}\right].
\end{equation}
This form of spectral density has previously been used to model an environment exhibiting a photonic band gap, where the localized gap in the density of states occurs for $\lambda_2W_1=\lambda_1W_2$ \cite{Garraway1997}. For the analysis below, we also note that the conditions $W_1\lambda_1>W_2\lambda_2$ ($\lambda_1>\lambda_2$) and $\lambda_2W_1\geq\lambda_1W_2$ are required since $D(\omega)$ is defined always to be positive, while $W_{1,2}$ must be normalized to $W_1-W_2=1$ [c.f. Eq. (\ref{eq:norm})].  \\
\indent Considering that Eq. (\ref{eq:PBG}) has poles located at positions $z_{1,2}=\xi-i\lambda_{1,2}$ in the lower-half complex $\omega$ plane, the two-time correlation function (\ref{eq:2tcorrE_res}) reads
\begin{equation}
	f(\tau) = \Omega^2_0\left(W_1e^{-iz_1\tau}-W_2e^{-iz_2\tau}\right),\quad\tau\geq0,
\end{equation}
with the discrete-mode couplings to the TLS given by $g'_1=\Omega_0\sqrt{W_1}$ and $g'_2=i\Omega_0\sqrt{W_2}$. Then, as $g'_2$ is clearly pure imaginary, one can apply the relevant formulas from Appendix \ref{appenC} to obtain the new frequency parameters for the TLS and discrete modes:
\begin{empheq}[left = \empheqlbrace]{align}\label{eq:params}
	\begin{split}
		&\tilde{g}'_1 = 0,\quad \tilde{g}'_2 = \Omega_0, \\
		&\tilde{\xi}_1 = \tilde{\xi}_2 = \xi,\\
		&V_{12} = \sqrt{W_1W_2}(\lambda_1-\lambda_2).
	\end{split}
\end{empheq}
This implies the TLS to now physically couple to the second auxiliary mode, which in turn interacts with the first mode. In an interaction frame generated by $H_0 = H_{S,0} + \xi\sum_m\tilde{b}^{\dagger}_m\tilde{b}_m$, the reduced Lindblad-type master equation for the enlarged system can subsequently be obtained by inserting Eqs. (\ref{eq:z_km}) and (\ref{eq:params}) into (\ref{eq:ME_eff}), where
\begin{align}\label{eq:PMME_BG}
	\frac{d}{dt}\rho(t)  &=-i[H_S(t),\rho(t)]\nonumber\\
				   &-i\Omega_0\left[e^{i\Delta t}\sigma_+\tilde{b}_2+e^{-i\Delta t}\tilde{b}^{\dagger}_2\sigma_-,\rho(t)\right]\nonumber\\
			            &-iV_{12}\left[\tilde{b}^{\dagger}_1\tilde{b}_2+\tilde{b}^{\dagger}_2\tilde{b}_1,\rho(t)\right]+\mathcal{D}[\rho(t)],
\end{align}
and 
\begin{align}
	\mathcal{D}[\rho] &= 2\sum^2_{m=1}\Gamma_m\left(\tilde{b}_m\rho \tilde{b}^{\dagger}_m-\frac{1}{2}\left\{\tilde{b}^{\dagger}_m\tilde{b}_m,\rho\right\}\right),\nonumber\\
	&\begin{cases}
		\Gamma_1 &= W_1\lambda_2-W_2\lambda_1 \\
		\Gamma_2 &= W_1\lambda_1-W_2\lambda_2.
	\end{cases}
\end{align}
As found previously with Eq. (\ref{eq:PMME}), this result generalizes the equivalent pseudomode master equation derived in \cite{Garraway1997} to now apply to a much greater variety of OQS models, including those \textit{not} limited to single-excitation regimes.

\section{Summary and outlook}\label{sec:6}

In conclusion, we have derived a master equation that provides a nonperturbative and non-Markovian description of an OQS dynamics within the RWA. This has been achieved by showing the reduced dynamics of an OQS to be indistinguishable under the effect of two different types of structured environment\textemdash one comprising an infinite collection of harmonic oscillators with a frequency-dependent coupling to the system and an auxiliary one comprising small number of discrete modes which are each coupled to an independent Markovian reservoir. The equivalence of these two models has subsequently been exploited to construct a general Lindblad master equation. In this way, the reduced dynamics of the original problem can be simulated efficiently within the framework of quantum trajectories, providing the Hilbert space dimension of the enlarged system is of a reasonable size. In particular, our approach has shown a full extension of the pseudomode theory to be possible to cases involving multiple excitations, where the strength of the OQS-discrete mode coupling and the degree of separation of timescales between the system and environment poses no restriction on the validity of the result. 

The procedure we have introduced to obtain the master equation relies on the connection between the poles of the spectral density contained in the complex $\omega$ plane, their residues, and the properties of the discrete modes used to represent the memory part of the environment. Specifically, the decay rates and couplings of the discrete modes are determined directly from the positions and residues of the poles of the spectral density function located in the lower-half complex plane. When the residues of these poles are evaluated to give complex coupling constants $g'_{jl}$, we have shown that our approach can still be applied using a non-Hermitian form of interaction Hamiltonian which reproduces the exact physical dynamics of the original open system. The master equation we obtain in this instance contains pathological terms that in general violate the positivity of the auxiliary system density matrix. For the two-discrete-mode case, these issues have been rectified via a change of basis of the discrete-mode terms to obtain a Lindblad-type master equation.

Our approach is particularly useful for analyzing non-Markovian effects in quantum optical systems where the physical conditions underlying the RWA are well satisfied. This includes experimental platforms relevant to cavity and circuit QED, in which such effects may be introduced into the dynamics by modifying the spectral density to be strongly frequency dependent. Moreover, since no particular assumptions are made on the open system itself, our results are also applicable to the study of many-body OQSs beyond Markovian interaction regimes \cite{Ribeiro2015,Xu2019}. The main practical advantage of our method in this context is that it allows for the reduced system dynamics to be simulated via a Lindblad master equation, which can be computed efficiently even for systems of relatively large size \cite{Daley2014}. 

The current treatment has been outlined for an initially empty reservoir, but a similar approach could be applied to a thermal (bosonic) environment at nonzero temperature. This would again involve evaluating the relevant two-time correlation functions of the environment in terms of the poles and residues of the spectral density $D(\omega)$ and utilizing the same connection between these poles and the discrete modes to perform the mapping. However, the mapping would now have to account for the poles contributed through the thermal part of the spectral density (depending on the Bose-Einstein distribution), of which there are infinitely many. At present the treatment is also limited to RWA-type interactions, based on the fact that the method we have used to evaluate the two-time correlation functions relies on the environment containing both positive- and negative-frequency components (i.e., where the spectral density function has support on the full real axis). Extending the pseudomode method beyond the RWA may be possible by combining our approach with the technique of Ref. \cite{Tamascelli2019}, where it is shown how to formally extend the environment definition to the negative-frequency domain under a less restrictive set of assumptions. 

Finally, an open problem relating to construction of the master equation is how to convert between the pathological and Lindblad form for when the OQS couples to more than two discrete modes with complex coefficients; as far as we are aware, no generalized form of transformation matrix allowing for the conversion has been determined beyond this case (although see, perhaps, the related inversion problem explored in Appendix B of Ref. \cite{Mascherpa2019}). Therefore future work in this area could focus on developing a systematic approach to regularizing the master equation for arbitrarily complicated environmental structures, which in itself would likely rely on a numerical implementation.

\section*{Acknowledgments}

G.P. thanks I. Sinayskiy for helpful discussions. This work is based upon research supported by the South African Research Chair Initiative, Grant No. 64812 of the Department of Science and Technology and the National Research Foundation of the Republic of South Africa.

\appendix

\section{Derivation of the discrete mode operator $b_l(t,s)$}\label{appenA}

As in the main text, we start by considering the time evolution of the Heisenberg picture operators $b_l(t,s)=U^{\dagger}_{MR}(t,s)b_lU_{MR}(t,s)$, whose corresponding equation of motion reads
\begin{align}\label{eq:HE_1}
	\frac{d}{dt}b_l(t,s)=-i\sqrt{\frac{\lambda_l}{\pi}}\int d\omega\,a_{Rl}(\omega,t,s)e^{-i(\omega-\xi_l)t}.
\end{align}
Simultaneously, the Heisenberg equation for the operators $a_{Rl}(\omega,t,s)=U^{\dagger}_{MR}(t,s)a_{Rl}(\omega)U_{MR}(t,s)$ is given by 
\begin{equation}\label{eq:HE_a}
	\frac{d}{dt}a_{Rl}(\omega,t,s)=-i\sqrt{\frac{\lambda_l}{\pi}}e^{i(\omega-\xi_l)t}b_l(t,s),
\end{equation}
which may be formally integrated to give
\begin{equation}\label{eq:a_R_sol}
	a_{Rl}(\omega,t,s) = a_{Rl}(\omega)-i\sqrt{\frac{\lambda_l}{\pi}}\int^t_{s}dt_1\,e^{i(\omega-\xi_l)t_1}b_l(t_1,s).
\end{equation}
Note that in this expression we have used the initial condition $a_{Rl}(\omega,s,s)\equiv a_{Rl}(\omega)$. Thus, inserting Eq. (\ref{eq:a_R_sol}) into (\ref{eq:HE_1}) yields the quantum Langevin equation
\begin{align}\label{eq:QLE_1}
	\frac{d}{dt}b_l(t,s)&=-i\sqrt{2\lambda_l}a^{\text{in}}_{Rl}(t)\nonumber\\
	&-\frac{\lambda_l}{\pi}\int^t_{s}dt_1\int d\omega\,b_l(t_1,s)e^{-i(\omega-\xi_l)(t-t_1)},
\end{align}	
where 
\begin{equation}\label{eq:a_in_appen}
	a^{\text{in}}_{Rl}(t)=\frac{1}{\sqrt{2\pi}}\int^{\infty}_{-\infty} d\omega\,a_{Rl}(\omega)e^{-i(\omega-\xi_l)t}
\end{equation}
defines the quantum noise operators of each of the reservoirs $R_l$. By now using the standard definition of a Dirac $\delta$ function
\begin{equation}\label{eq:delta_func}
	\int^{\infty}_{-\infty}d\omega\,e^{-i\omega(t-t_1)}=2\pi\delta(t-t_1)
\end{equation}
and the integral property 
\begin{equation}
	2\lambda_l\int^t_{s}dt_1\delta(t-t_1)b_l(t_1,s) = \lambda_lb_l(t,s),
\end{equation}
the quantum Langevin equation (\ref{eq:QLE_1}) can be expressed as
\begin{equation}
	\frac{d}{dt}b_l(t,s) = -\lambda_lb_l(t,s) - i\sqrt{2\lambda_l}a^{\text{in}}_{Rl}(t).
\end{equation}
Since this is nothing but an ordinary inhomogeneous differential equation for $b_l(t,s)$, its solution may be written in the form (see, for example, the variation of parameters approach in Ref. \cite{Chicone2006})
\begin{equation}\label{eq:b_l(t)_appen}
	b_l(t,s) = e^{-\lambda_l(t-s)}b_l(0)-i\sqrt{2\lambda_l}\int^t_{s}dt_1e^{-\lambda_l(t-t_1)}a^{\text{in}}_{Rl}(t_1),
\end{equation}
as given in Eq. (\ref{eq:b_l(t)}) of the main text.

\section{Equivalence between the reduced density operators $\rho_S(t)$ and $\rho'_S(t)$}\label{appenB}

Here we outline the proof showing that the reduced system dynamics of $\rho_S(t)$ and $\rho'_S(t)$ will be equivalent as long as the two-time correlation functions (\ref{eq:2tcorrE}) and (\ref{eq:2tcorrE'}) share the same time dependence, even when the interaction Hamiltonian may be non-Hermitian. To this end, let us first restate our main assumptions of factorizing initial conditions $\rho_{SE}(0)=\rho_S(0)\otimes\rho_E(0)$ [$\rho_{SMR}(0)=\rho_S(0)\otimes\rho_{MR}(0)$], where $\rho_E(0)$ [$\rho_{MR}(0)$] is restricted to the vacuum state (\ref{eq:rho_E}) [(\ref{eq:rho_MR})]. We may then proceed by noting that a general solution to the reduced density operator $\rho_S(t)$ [$\rho'_S(t)$] can be written as an expansion over the noise operators in Eq. (\ref{eq:H_I_int}) and (\ref{eq:H_I'_int}) \cite{Tamascelli2018,Lambert2019}. Indeed, within an interaction picture generated by the unitary transformation (\ref{eq:U_0}), the evolution of the density operator $\rho_{SE}(t)$ is determined by the von Neumann equation
\begin{align}
	\frac{d}{dt}\rho_{SE}(t) &= -i[H_S(t)+H_I(t),\rho_{SE}(t)]\nonumber\\
					   &\equiv \mathcal{L}(t)\rho_{SE}(t),
\end{align}
which can immediately be solved to obtain
\begin{equation}\label{eq:rho_SE}
	\rho_{SE}(t) = \mathcal{T}\text{exp}\left[\int^t_0ds\,\mathcal{L}(s)\right]\rho_{SE}(0).
\end{equation}
If we now expand the Dyson series of Eq. (\ref{eq:rho_SE}) and trace out the environmental degrees of freedom to obtain $\rho_S(t) = \text{Tr}_E[\rho_{SE}(t)]$, the reduced system density operator can be derived in the form 
\begin{widetext}
\begin{align}\label{eq:rho_S_exp}
	\rho_S(t) &= \left(1+\sum^{\infty}_{n=1}\int^t_0dt_1\cdots\int^{t_{n-1}}_0dt_n\left\langle\mathcal{L}(t_1)\cdots\mathcal{L}(t_n)\right\rangle_E\right)\rho_S(0)\nonumber\\
		       &\equiv \Phi(t,0)\rho_S(0),
\end{align}
\end{widetext}
where $\langle \mathcal{L}(t_1)\mathcal{L}(t_2)\cdots\mathcal{L}(t_n)\rangle_E$ ($t_1\geq t_2\geq ...$) contains the $n^{\text{th}}$ time-ordered moments of the environmental noise operators $B_j(t)$, $B^{\dagger}_j(t)$ [and system coupling operators $c_j(t)$, $c^{\dagger}_j(t)$], as well as terms connected to the free system Hamiltonian $H_S(t)$. Similarly, for $\rho'_S(t)$ we may also introduce the von Neumann equation
\begin{align}\label{eq:rho_SMR}
	\frac{d}{dt}\rho_{SMR}(t) &= -i[H_S(t) + H'_I(t),\rho_{SMR}(t)]\nonumber\\
					    &\equiv \mathcal{L}'(t)\rho_{SMR}(t)
\end{align}
and follow an analogous procedure to obtain 
\begin{widetext}
\begin{align}\label{eq:rho'_S_exp}
	\rho'_S(t) &= \left(1+\sum^{\infty}_{n=1}\int^t_0dt_1\cdots\int^{t_{n-1}}_0dt_n\left\langle\mathcal{L}'(t_1)\cdots\mathcal{L}'(t_n)\right\rangle_{E'}\right)\rho_S(0)\nonumber\\
		       &\equiv \Phi'(t,0)\rho_S(0).
\end{align}
\end{widetext}
Note that in cases when the Hamiltonian is non-Hermitian, Eq. (\ref{eq:rho_SMR}) defines an $S+E'$ evolution which has been modified to be nonunitary, i.e., we are intentionally \textit{not} taking the Hermitian conjugate of $H'_I(t)$ when acting on the right of $\rho_{SMR}(t)$ \cite{Lambert2019}. The reason for this is to ensure the two expansions of $\rho_S(t)$ and $\rho'_S(t)$ share the same structure. As a result, it can be observed in the general case that the dynamics of the reduced density operators will be equivalent as long as first-, second-, etc., order terms in the two expansions share an equal time dependence. \\
\indent Now, since the free system dynamics is unchanged between environment configurations, as well as the coupling operators $c_j(t)$ ($c^{\dagger}_j(t)$), we see that the only differences between $\rho_S(t)$ and $\rho'_S(t)$ will be contained in the $n^{\text{th}}$-order moments of the corresponding noise operators. However, taking into account that the states $\rho_E(0)$ and $\rho_{MR}(0)$ are Gaussian, Wick's theorem implies the complete set of moments appearing in either expansion to factorize into first- and second-order moments, so that from $\text{Tr}_E[B_j(t)\rho_E(0)]=0=\text{Tr}_{MR}[B'_j(t)\rho_{MR}(0)]$, if we also have
\begin{equation}
	f_{jk}(t-s) = \langle B'_j(t)B'^{\,\dagger}_k(s)\rangle_{E'},\quad \forall j,k,\,t\geq s,
\end{equation}
then the reduced expansions (\ref{eq:rho_S_exp}) and (\ref{eq:rho'_S_exp}) will be indistinguishable in the sense that
\begin{equation}\label{eq:equiv_appen}
	\Phi(t,0)=\Phi'(t,0)\implies \rho_S(t)=\rho'_S(t).
\end{equation}
As a final remark, we note that the above proof implies the Heisenberg equations of motion for operators defined in the auxiliary model to adopt the same form as Eq. (\ref{eq:rho_SMR}) in cases where $H'_I(t)$ is non-Hermitian [c.f. Eqs. (\ref{eq:HE_A})-(\ref{eq:aveHE_2})]. This leads to a master equation which is equivalent to Eq. (\ref{eq:ME}), but which for complex OQS-discrete-mode couplings is of a pathological (non-Lindblad) form.

\section{Analytical expressions for $\tilde{z}_{mm'}$ ($\tilde{\xi}_m$, $\Gamma_m$, $V_{12}$) and $\tilde{g}'_{jm}$}\label{appenC}

In this Appendix we derive exact analytical expressions for the coefficients $\tilde{z}_{mm'}$ and couplings parameter $\tilde{g}'_{jm}$ in cases involving two discrete modes. To do so we recall the treatment of essentially the same problem considered by Garraway in Sec. VB of \cite{Garraway1997}, where in that work a procedure was implemented to convert between pathological and Lindblad forms of the master equation with equivalent coefficients to Eqs. (\ref{eq:ME_eff_1}) and (\ref{eq:ME_eff}); here we then utilize this very same procedure to obtain a more general result overall. \\
\indent Considering first that the transformation matrix $U_{ml}$ for $m,l=1,2$ can generally be written in the form of a complex rotation about an angle $\theta_0$,
\begin{equation}
	{\bm U}(\theta_0) = 
	\begin{pmatrix}
		\cos\theta_0 & \sin\theta_0 \\
		-\sin\theta_0 & \cos\theta_0	
	\end{pmatrix},
\end{equation}
the similarity transformation in Eq. (\ref{eq:sim_trans_1}) can be evaluated explicitly to yield 
\begin{align}\label{eq:z_km_appen}
	\tilde{z}_{mm'} &= \frac{1}{2}\big(z_1+z_2+(-1)^m\,\Delta z\cos2\theta_0\big)\delta_{mm'}\nonumber\\
	 		     &+ \frac{\Delta z\sin2\theta_0}{2}\big(\delta_{mm'-1}+\delta_{mm'+1}\big),\quad m,m'=1,2,
\end{align}
where 
\begin{equation}
	\Delta z=z_2-z_1\equiv|\Delta z|\text{exp}(i\theta_z)
\end{equation}
defines the distance between the two poles of $D(\omega)$ in the lower-half complex plane. In turn, we may also proceed to parametrize the OQS-discrete-mode couplings $g'_{jl}=\Omega_j\sqrt{-ir_l}$ via
\begin{equation}\label{eq:g_vec}
	{\bm g}'_j = \Omega_j
	\begin{pmatrix}
		\cos\theta_1 \\
		\sin\theta_1	
	\end{pmatrix}
	\equiv \Omega_j{\bm U}(-\theta_1)
	\begin{pmatrix}
		1 \\
		0	
	\end{pmatrix},
\end{equation}
which by construction satisfies the normalization property (\ref{eq:norm_g}), i.e., $\|{\bm g}'_j\|^2=\Omega^2_j$. Since the relative magnitude of these couplings depends only on a single free parameter $\theta_1$, we can more conveniently characterize Eq. (\ref{eq:g_vec}) in terms a complex ratio $\mu$, where
\begin{equation}
	\mu=\tan\theta_1 = g'_{j2}/g'_{j1}.
\end{equation}
We will now proceed to write the new couplings $\tilde{g}'_{jm}$ as
\begin{align}
	\tilde{\bm g}'_j &= {\bm U}(\theta_0){\bm g}'_j\nonumber\\
		               &= \Omega_j{\bm U}(\theta_0-\theta_1)
	\begin{pmatrix}
		1 \\
		0	
	\end{pmatrix},
\end{align}
so that with the help of these basic definitions, the rotation angle $\theta_0$ can be determined in principle according to the physical constraints placed on the quantities $\tilde{g}'_{jm}$ and $\Gamma_{mm'}=-\text{Im}[\tilde{z}_{mm'}]$ (see the details provided in Sec. \ref{sec:5A} of the main text). This is in part what we shall now discuss below. \\
\indent Following Ref. \cite{Garraway1997}, we first note that the off-diagonal elements of Eq. (\ref{eq:z_km_appen}) are constrained to be real. As such, we may write
\begin{equation}
	\tilde{z}_{mm'} \equiv (\tilde{\xi}_m-i\Gamma_m)\delta_{mm'}+V_{12}\left(\delta_{mm'-1}+\delta_{mm'+1}\right)
\end{equation}
with 
\begin{empheq}[left = \empheqlbrace]{align}\label{eq:params_appen}
	\begin{split}
		&\tilde{\xi}_m = \frac{1}{2}\big(\xi_1+\xi_2+(-1)^m\,\text{Re}\left[\Delta z\cos2\theta_0\right]\big)\\
		&\Gamma_m = \frac{1}{2}\big(\lambda_1+\lambda_2+(-1)^{m-1}\,\text{Im}\left[\Delta z\cos2\theta_0\right]\big)\\
		&V_{12} = \frac{\Delta z\sin2\theta_0}{2}.
	\end{split}
\end{empheq}
While this is not a necessary condition for the approach, due to the fact that the resulting matrix $\Gamma_{mm'}$ is diagonal, i.e., $\Gamma_{mm'}=\Gamma_m\delta_{mm'}$, its positivity can now be guaranteed by the simpler constraint that only $\Gamma_m$ must be non-negative ($\Gamma_m\geq 0$):
\begin{align}\label{eq:constraint_1}
	\lambda_1+\lambda_2&\geq (-1)^m\big\{(\xi_1-\xi_2)\text{Im}[\cos2\theta_0]\nonumber\\
					  &+(\lambda_1-\lambda_2)\text{Re}[\cos2\theta_0]\big\},\quad m=1,2.
\end{align}
This inequality, which relates the allowed values $\theta_0$ to the parameters of correlation function (\ref{eq:2tcorrE_res}), may or may not be possible to satisfy depending on the choice $\lambda_l$ or $\xi_l$, so that its validity must be checked in the general case (though this should not be too demanding to satisfy based on the examples shown in Ref. \cite{Garraway1997}). The second constraint comes from the requirement that $\tilde{g}'_{jm}$ must be real, and so
\begin{equation}\label{eq:constraint_2}
	\text{Im}[\theta_0] = \text{Im}[\theta_1].
\end{equation}
Thus the problem of determining $\tilde{z}_{mm'}$ and $\tilde{g}'_{jm}$ reduces to eliminating $\theta_0$ from Eq. (\ref{eq:params_appen}) through the constraints (\ref{eq:constraint_1})-(\ref{eq:constraint_2}). Although this will not be explicitly shown here, it ultimately follows from Ref. \cite{Garraway1997} that $\tilde{z}_{mm'}(\tilde{\xi}_m,\Gamma_m,V_{12})$ and $\tilde{g}'_{jm}$ can be expressed solely as functions of the known variables $\Delta z$, $\mu$, and $\theta_z$. Indeed, for the coupling between discrete modes $V_{12}$, we find
\begin{equation}
	V_{12}(\mu,\theta_z) = -\frac{|\Delta z|}{|1+\mu^2|}\frac{\text{Im}[\mu](1+|\mu|^2)}{\sqrt{\sin^2\theta_z(1+|\mu|^2)^2+(2\,\text{Im}[\mu]\cos\theta_z)^2}},
\end{equation}
while the discrete-mode frequencies and decay rates $\tilde{\xi}_m$ and $\Gamma_m$ are given by ($k=1,2$)
\onecolumngrid
\begin{align}
	\tilde{\xi}_m(\mu,\theta_z)&=\frac{1}{2}\left(\xi_1+\xi_2+(-1)^m\,\text{Re}\left[\frac{-\Delta z}{\left|1+\mu^2\right|}\frac{(1+|\mu|^2)^2\sin\theta_z+i(2\,\text{Im}[\mu])^2\cos\theta_z}{\sqrt{(1+|\mu|^2)^2\sin^2\theta_z+(2\,\text{Im}[\mu]\text{cos}\theta_z)^2}}\right]\right),\\
	\Gamma_m(\mu,\theta_z)&=\frac{1}{2}\left(\lambda_1+\lambda_2+(-1)^{m-1}\,\text{Im}\left[\frac{-\Delta z}{\left|1+\mu^2\right|}\frac{(1+|\mu|^2)^2\sin\theta_z+i(2\,\text{Im}[\mu])^2\cos\theta_z}{\sqrt{(1+|\mu|^2)^2\sin^2\theta_z+(2\,\text{Im}[\mu]\text{cos}\theta_z)^2}}\right]\right).
\end{align}
Finally, the expressions for the OQS-discrete-mode couplings read
\begin{align}
	(\tilde{g}'_{j1})^2 &= \frac{\Omega^2_j}{2}\left[1-\frac{(1-|\mu|^4)\sin\theta_z-4\,\text{Re}[\mu]\text{Im}[\mu]\cos\theta_z}{\left|1+\mu^2\right|\sqrt{(1+|\mu|^2)^2\sin^2\theta_z+(2\text{Im}[\mu]\cos\theta_z)^2}}\right], \\
	(\tilde{g}'_{j2})^2 & = \frac{\Omega^2_j}{2}\left[1+\frac{(1-|\mu|^4)\sin\theta_z-4\,\text{Re}[\mu]\text{Im}[\mu]\cos\theta_z}{\left|1+\mu^2\right|\sqrt{(1+|\mu|^2)^2\sin^2\theta_z+(2\text{Im}[\mu]\cos\theta_z)^2}}\right].
\end{align}

\twocolumngrid

\bibliographystyle{apsrev4-1} 

\bibliography{PM_references}

\begin{thebibliography}{64}%
\makeatletter
\providecommand \@ifxundefined [1]{%
 \@ifx{#1\undefined}
}%
\providecommand \@ifnum [1]{%
 \ifnum #1\expandafter \@firstoftwo
 \else \expandafter \@secondoftwo
 \fi
}%
\providecommand \@ifx [1]{%
 \ifx #1\expandafter \@firstoftwo
 \else \expandafter \@secondoftwo
 \fi
}%
\providecommand \natexlab [1]{#1}%
\providecommand \enquote  [1]{``#1''}%
\providecommand \bibnamefont  [1]{#1}%
\providecommand \bibfnamefont [1]{#1}%
\providecommand \citenamefont [1]{#1}%
\providecommand \href@noop [0]{\@secondoftwo}%
\providecommand \href [0]{\begingroup \@sanitize@url \@href}%
\providecommand \@href[1]{\@@startlink{#1}\@@href}%
\providecommand \@@href[1]{\endgroup#1\@@endlink}%
\providecommand \@sanitize@url [0]{\catcode `\\12\catcode `\$12\catcode
  `\&12\catcode `\#12\catcode `\^12\catcode `\_12\catcode `\%12\relax}%
\providecommand \@@startlink[1]{}%
\providecommand \@@endlink[0]{}%
\providecommand \url  [0]{\begingroup\@sanitize@url \@url }%
\providecommand \@url [1]{\endgroup\@href {#1}{\urlprefix }}%
\providecommand \urlprefix  [0]{URL }%
\providecommand \Eprint [0]{\href }%
\providecommand \doibase [0]{http://dx.doi.org/}%
\providecommand \selectlanguage [0]{\@gobble}%
\providecommand \bibinfo  [0]{\@secondoftwo}%
\providecommand \bibfield  [0]{\@secondoftwo}%
\providecommand \translation [1]{[#1]}%
\providecommand \BibitemOpen [0]{}%
\providecommand \bibitemStop [0]{}%
\providecommand \bibitemNoStop [0]{.\EOS\space}%
\providecommand \EOS [0]{\spacefactor3000\relax}%
\providecommand \BibitemShut  [1]{\csname bibitem#1\endcsname}%
\let\auto@bib@innerbib\@empty
%</preamble>
\bibitem [{\citenamefont {Breuer}\ and\ \citenamefont
  {Petruccione}(2002)}]{BreuerTOQS2002}%
  \BibitemOpen
  \bibfield  {author} {\bibinfo {author} {\bibfnamefont {H.-P.}\ \bibnamefont
  {Breuer}}\ and\ \bibinfo {author} {\bibfnamefont {F.}~\bibnamefont
  {Petruccione}},\ }\href@noop {} {\emph {\bibinfo {title} {The theory of open
  quantum systems}}}\ (\bibinfo  {publisher} {Oxford University Press},\
  \bibinfo {address} {New York},\ \bibinfo {year} {2002})\BibitemShut {NoStop}%
\bibitem [{\citenamefont {Gardiner}\ and\ \citenamefont
  {Zoller}(2005)}]{Gardiner2005}%
  \BibitemOpen
  \bibfield  {author} {\bibinfo {author} {\bibfnamefont {C.~W.}\ \bibnamefont
  {Gardiner}}\ and\ \bibinfo {author} {\bibfnamefont {P.}~\bibnamefont
  {Zoller}},\ }\href@noop {} {\emph {\bibinfo {title} {Quantum Noise}}}\
  (\bibinfo  {publisher} {Springer},\ \bibinfo {address} {Berlin},\ \bibinfo
  {year} {2005})\BibitemShut {NoStop}%
\bibitem [{\citenamefont {Nielson}\ and\ \citenamefont
  {Chuang}(2010)}]{Nielson2010}%
  \BibitemOpen
  \bibfield  {author} {\bibinfo {author} {\bibfnamefont {M.~A.}\ \bibnamefont
  {Nielson}}\ and\ \bibinfo {author} {\bibfnamefont {I.~L.}\ \bibnamefont
  {Chuang}},\ }\href@noop {} {\emph {\bibinfo {title} {Quantum Computation and
  Quantum Information}}}\ (\bibinfo  {publisher} {Cambridge University Press},\
  \bibinfo {address} {Cambridge, England},\ \bibinfo {year} {2010})\BibitemShut
  {NoStop}%
\bibitem [{\citenamefont {Schlosshauer}(2019)}]{Schlosshauer2019}%
  \BibitemOpen
  \bibfield  {author} {\bibinfo {author} {\bibfnamefont {M.}~\bibnamefont
  {Schlosshauer}},\ }\href {\doibase 10.1016/j.physrep.2019.10.001} {\bibfield
  {journal} {\bibinfo  {journal} {Phys. Rep.}\ }\textbf {\bibinfo {volume}
  {831}},\ \bibinfo {pages} {1} (\bibinfo {year} {2019})}\BibitemShut {NoStop}%
\bibitem [{\citenamefont {Zurek}(1991)}]{Zurek1991}%
  \BibitemOpen
  \bibfield  {author} {\bibinfo {author} {\bibfnamefont {W.~H.}\ \bibnamefont
  {Zurek}},\ }\href {\doibase 10.1063/1.881293} {\bibfield  {journal} {\bibinfo
   {journal} {Physics Today}\ }\textbf {\bibinfo {volume} {44}},\ \bibinfo
  {pages} {36} (\bibinfo {year} {1991})}\BibitemShut {NoStop}%
\bibitem [{\citenamefont {Carmichael}(1993)}]{Carmichael1993}%
  \BibitemOpen
  \bibfield  {author} {\bibinfo {author} {\bibfnamefont {H.~J.}\ \bibnamefont
  {Carmichael}},\ }\href@noop {} {\emph {\bibinfo {title} {An Open Systems
  Approach to Quantum Optics}}}\ (\bibinfo  {publisher} {Springer-Verlag},\
  \bibinfo {address} {Berlin},\ \bibinfo {year} {1993})\BibitemShut {NoStop}%
\bibitem [{\citenamefont {Leggett}\ \emph {et~al.}(1987)\citenamefont
  {Leggett}, \citenamefont {Chakravarty}, \citenamefont {Dorsey}, \citenamefont
  {Fisher}, \citenamefont {Garg},\ and\ \citenamefont {Zwerger}}]{Leggett1987}%
  \BibitemOpen
  \bibfield  {author} {\bibinfo {author} {\bibfnamefont {A.~J.}\ \bibnamefont
  {Leggett}}, \bibinfo {author} {\bibfnamefont {S.}~\bibnamefont
  {Chakravarty}}, \bibinfo {author} {\bibfnamefont {A.~T.}\ \bibnamefont
  {Dorsey}}, \bibinfo {author} {\bibfnamefont {M.~P.~A.}\ \bibnamefont
  {Fisher}}, \bibinfo {author} {\bibfnamefont {A.}~\bibnamefont {Garg}}, \ and\
  \bibinfo {author} {\bibfnamefont {W.}~\bibnamefont {Zwerger}},\ }\href
  {\doibase 10.1103/RevModPhys.59.1} {\bibfield  {journal} {\bibinfo  {journal}
  {Rev. Mod. Phys.}\ }\textbf {\bibinfo {volume} {59}},\ \bibinfo {pages} {1}
  (\bibinfo {year} {1987})}\BibitemShut {NoStop}%
\bibitem [{\citenamefont {Vinjanampathy}\ and\ \citenamefont
  {Anders}(2016)}]{Vinjanampathy2016}%
  \BibitemOpen
  \bibfield  {author} {\bibinfo {author} {\bibfnamefont {S.}~\bibnamefont
  {Vinjanampathy}}\ and\ \bibinfo {author} {\bibfnamefont {J.}~\bibnamefont
  {Anders}},\ }\href {\doibase 10.1080/00107514.2016.1201896} {\bibfield
  {journal} {\bibinfo  {journal} {Contemp. Phys.}\ }\textbf {\bibinfo {volume}
  {57}},\ \bibinfo {pages} {545} (\bibinfo {year} {2016})}\BibitemShut
  {NoStop}%
\bibitem [{\citenamefont {Gonz\'alez-Tudela}\ and\ \citenamefont
  {Porras}(2013)}]{Tudela2013}%
  \BibitemOpen
  \bibfield  {author} {\bibinfo {author} {\bibfnamefont {A.}~\bibnamefont
  {Gonz\'alez-Tudela}}\ and\ \bibinfo {author} {\bibfnamefont {D.}~\bibnamefont
  {Porras}},\ }\href {\doibase 10.1103/PhysRevLett.110.080502} {\bibfield
  {journal} {\bibinfo  {journal} {Phys. Rev. Lett.}\ }\textbf {\bibinfo
  {volume} {110}},\ \bibinfo {pages} {080502} (\bibinfo {year}
  {2013})}\BibitemShut {NoStop}%
\bibitem [{\citenamefont {Brandes}(2005)}]{Brandes2005}%
  \BibitemOpen
  \bibfield  {author} {\bibinfo {author} {\bibfnamefont {T.}~\bibnamefont
  {Brandes}},\ }\href {\doibase 10.1016/j.physrep.2004.12.002} {\bibfield
  {journal} {\bibinfo  {journal} {Phys. Rep.}\ }\textbf {\bibinfo {volume}
  {408}},\ \bibinfo {pages} {315} (\bibinfo {year} {2005})}\BibitemShut
  {NoStop}%
\bibitem [{\citenamefont {Breuer}\ \emph {et~al.}(2016)\citenamefont {Breuer},
  \citenamefont {Laine}, \citenamefont {Piilo},\ and\ \citenamefont
  {Vacchini}}]{Breuer2016}%
  \BibitemOpen
  \bibfield  {author} {\bibinfo {author} {\bibfnamefont {H.-P.}\ \bibnamefont
  {Breuer}}, \bibinfo {author} {\bibfnamefont {E.-M.}\ \bibnamefont {Laine}},
  \bibinfo {author} {\bibfnamefont {J.}~\bibnamefont {Piilo}}, \ and\ \bibinfo
  {author} {\bibfnamefont {B.}~\bibnamefont {Vacchini}},\ }\href {\doibase
  10.1103/RevModPhys.88.021002} {\bibfield  {journal} {\bibinfo  {journal}
  {Rev. Mod. Phys.}\ }\textbf {\bibinfo {volume} {88}},\ \bibinfo {pages}
  {021002} (\bibinfo {year} {2016})}\BibitemShut {NoStop}%
\bibitem [{\citenamefont {de~Vega}\ and\ \citenamefont
  {Alonso}(2017)}]{DeVega2017}%
  \BibitemOpen
  \bibfield  {author} {\bibinfo {author} {\bibfnamefont {I.}~\bibnamefont
  {de~Vega}}\ and\ \bibinfo {author} {\bibfnamefont {D.}~\bibnamefont
  {Alonso}},\ }\href {\doibase 10.1103/RevModPhys.89.015001} {\bibfield
  {journal} {\bibinfo  {journal} {Rev. Mod. Phys.}\ }\textbf {\bibinfo {volume}
  {89}},\ \bibinfo {pages} {015001} (\bibinfo {year} {2017})}\BibitemShut
  {NoStop}%
\bibitem [{\citenamefont {Li}\ \emph {et~al.}(2018)\citenamefont {Li},
  \citenamefont {Hall},\ and\ \citenamefont {Wiseman}}]{Li2018}%
  \BibitemOpen
  \bibfield  {author} {\bibinfo {author} {\bibfnamefont {L.}~\bibnamefont
  {Li}}, \bibinfo {author} {\bibfnamefont {M.~J.~W.}\ \bibnamefont {Hall}}, \
  and\ \bibinfo {author} {\bibfnamefont {H.~M.}\ \bibnamefont {Wiseman}},\
  }\href {\doibase 10.1016/j.physrep.2018.07.001} {\bibfield  {journal}
  {\bibinfo  {journal} {Phys. Rep.}\ }\textbf {\bibinfo {volume} {759}},\
  \bibinfo {pages} {1} (\bibinfo {year} {2018})}\BibitemShut {NoStop}%
\bibitem [{\citenamefont {Lemmer}\ \emph {et~al.}(2018)\citenamefont {Lemmer},
  \citenamefont {Cormick}, \citenamefont {Tamascelli}, \citenamefont {Schaetz},
  \citenamefont {Huelga},\ and\ \citenamefont {Plenio}}]{Lemmer2018}%
  \BibitemOpen
  \bibfield  {author} {\bibinfo {author} {\bibfnamefont {A.}~\bibnamefont
  {Lemmer}}, \bibinfo {author} {\bibfnamefont {C.}~\bibnamefont {Cormick}},
  \bibinfo {author} {\bibfnamefont {D.}~\bibnamefont {Tamascelli}}, \bibinfo
  {author} {\bibfnamefont {T.}~\bibnamefont {Schaetz}}, \bibinfo {author}
  {\bibfnamefont {S.~F.}\ \bibnamefont {Huelga}}, \ and\ \bibinfo {author}
  {\bibfnamefont {M.~B.}\ \bibnamefont {Plenio}},\ }\href {\doibase
  10.1088/1367-2630/aac87d} {\bibfield  {journal} {\bibinfo  {journal} {New J.
  Phys.}\ }\textbf {\bibinfo {volume} {20}},\ \bibinfo {pages} {073002}
  (\bibinfo {year} {2018})}\BibitemShut {NoStop}%
\bibitem [{\citenamefont {Ribeiro}\ and\ \citenamefont
  {Vieira}(2015)}]{Ribeiro2015}%
  \BibitemOpen
  \bibfield  {author} {\bibinfo {author} {\bibfnamefont {P.}~\bibnamefont
  {Ribeiro}}\ and\ \bibinfo {author} {\bibfnamefont {V.~R.}\ \bibnamefont
  {Vieira}},\ }\href {\doibase 10.1103/PhysRevB.92.100302} {\bibfield
  {journal} {\bibinfo  {journal} {Phys. Rev. B}\ }\textbf {\bibinfo {volume}
  {92}},\ \bibinfo {pages} {100302(R)} (\bibinfo {year} {2015})}\BibitemShut
  {NoStop}%
\bibitem [{\citenamefont {Hoeppe}\ \emph {et~al.}(2012)\citenamefont {Hoeppe},
  \citenamefont {Wolff}, \citenamefont {K\"uchenmeister}, \citenamefont
  {Niegemann}, \citenamefont {Drescher}, \citenamefont {Benner},\ and\
  \citenamefont {Busch}}]{Hoeppe2012}%
  \BibitemOpen
  \bibfield  {author} {\bibinfo {author} {\bibfnamefont {U.}~\bibnamefont
  {Hoeppe}}, \bibinfo {author} {\bibfnamefont {C.}~\bibnamefont {Wolff}},
  \bibinfo {author} {\bibfnamefont {J.}~\bibnamefont {K\"uchenmeister}},
  \bibinfo {author} {\bibfnamefont {J.}~\bibnamefont {Niegemann}}, \bibinfo
  {author} {\bibfnamefont {M.}~\bibnamefont {Drescher}}, \bibinfo {author}
  {\bibfnamefont {H.}~\bibnamefont {Benner}}, \ and\ \bibinfo {author}
  {\bibfnamefont {K.}~\bibnamefont {Busch}},\ }\href {\doibase
  10.1103/PhysRevLett.108.043603} {\bibfield  {journal} {\bibinfo  {journal}
  {Phys. Rev. Lett.}\ }\textbf {\bibinfo {volume} {108}},\ \bibinfo {pages}
  {043603} (\bibinfo {year} {2012})}\BibitemShut {NoStop}%
\bibitem [{\citenamefont {Haase}\ \emph {et~al.}(2018)\citenamefont {Haase},
  \citenamefont {Vetter}, \citenamefont {Unden}, \citenamefont {Smirne},
  \citenamefont {Rosskopf}, \citenamefont {Naydenov}, \citenamefont {Stacey},
  \citenamefont {Jelezko}, \citenamefont {Plenio},\ and\ \citenamefont
  {Huelga}}]{Haase2018}%
  \BibitemOpen
  \bibfield  {author} {\bibinfo {author} {\bibfnamefont {J.~F.}\ \bibnamefont
  {Haase}}, \bibinfo {author} {\bibfnamefont {P.~J.}\ \bibnamefont {Vetter}},
  \bibinfo {author} {\bibfnamefont {T.}~\bibnamefont {Unden}}, \bibinfo
  {author} {\bibfnamefont {A.}~\bibnamefont {Smirne}}, \bibinfo {author}
  {\bibfnamefont {J.}~\bibnamefont {Rosskopf}}, \bibinfo {author}
  {\bibfnamefont {B.}~\bibnamefont {Naydenov}}, \bibinfo {author}
  {\bibfnamefont {A.}~\bibnamefont {Stacey}}, \bibinfo {author} {\bibfnamefont
  {F.}~\bibnamefont {Jelezko}}, \bibinfo {author} {\bibfnamefont {M.~B.}\
  \bibnamefont {Plenio}}, \ and\ \bibinfo {author} {\bibfnamefont {S.~F.}\
  \bibnamefont {Huelga}},\ }\href {\doibase 10.1103/PhysRevLett.121.060401}
  {\bibfield  {journal} {\bibinfo  {journal} {Phys. Rev. Lett.}\ }\textbf
  {\bibinfo {volume} {121}},\ \bibinfo {pages} {060401} (\bibinfo {year}
  {2018})}\BibitemShut {NoStop}%
\bibitem [{\citenamefont {Wang}\ \emph {et~al.}(2018)\citenamefont {Wang},
  \citenamefont {Hou}, \citenamefont {Huang}, \citenamefont {Zhang},
  \citenamefont {Ouyang}, \citenamefont {Wang}, \citenamefont {Huang},
  \citenamefont {Zhang}, \citenamefont {He}, \citenamefont {Chang},\ and\
  \citenamefont {Duan}}]{Wang2018}%
  \BibitemOpen
  \bibfield  {author} {\bibinfo {author} {\bibfnamefont {F.}~\bibnamefont
  {Wang}}, \bibinfo {author} {\bibfnamefont {P.~Y.}\ \bibnamefont {Hou}},
  \bibinfo {author} {\bibfnamefont {Y.~Y.}\ \bibnamefont {Huang}}, \bibinfo
  {author} {\bibfnamefont {W.~G.}\ \bibnamefont {Zhang}}, \bibinfo {author}
  {\bibfnamefont {X.~L.}\ \bibnamefont {Ouyang}}, \bibinfo {author}
  {\bibfnamefont {X.}~\bibnamefont {Wang}}, \bibinfo {author} {\bibfnamefont
  {X.~Z.}\ \bibnamefont {Huang}}, \bibinfo {author} {\bibfnamefont {H.~L.}\
  \bibnamefont {Zhang}}, \bibinfo {author} {\bibfnamefont {L.}~\bibnamefont
  {He}}, \bibinfo {author} {\bibfnamefont {X.~Y.}\ \bibnamefont {Chang}}, \
  and\ \bibinfo {author} {\bibfnamefont {L.~M.}\ \bibnamefont {Duan}},\ }\href
  {\doibase 10.1103/PhysRevB.98.064306} {\bibfield  {journal} {\bibinfo
  {journal} {Phys. Rev. B}\ }\textbf {\bibinfo {volume} {98}},\ \bibinfo
  {pages} {064306} (\bibinfo {year} {2018})}\BibitemShut {NoStop}%
\bibitem [{\citenamefont {Peng}\ \emph {et~al.}(2018)\citenamefont {Peng},
  \citenamefont {Xu}, \citenamefont {Xu}, \citenamefont {Huang}, \citenamefont
  {Wang}, \citenamefont {Kong}, \citenamefont {Rong}, \citenamefont {Shi},
  \citenamefont {Duan},\ and\ \citenamefont {Du}}]{Peng2018}%
  \BibitemOpen
  \bibfield  {author} {\bibinfo {author} {\bibfnamefont {S.}~\bibnamefont
  {Peng}}, \bibinfo {author} {\bibfnamefont {X.}~\bibnamefont {Xu}}, \bibinfo
  {author} {\bibfnamefont {K.}~\bibnamefont {Xu}}, \bibinfo {author}
  {\bibfnamefont {P.}~\bibnamefont {Huang}}, \bibinfo {author} {\bibfnamefont
  {P.}~\bibnamefont {Wang}}, \bibinfo {author} {\bibfnamefont {X.}~\bibnamefont
  {Kong}}, \bibinfo {author} {\bibfnamefont {X.}~\bibnamefont {Rong}}, \bibinfo
  {author} {\bibfnamefont {F.}~\bibnamefont {Shi}}, \bibinfo {author}
  {\bibfnamefont {C.}~\bibnamefont {Duan}}, \ and\ \bibinfo {author}
  {\bibfnamefont {J.}~\bibnamefont {Du}},\ }\href {\doibase
  10.1016/j.scib.2018.02.017} {\bibfield  {journal} {\bibinfo  {journal} {Sci.
  Bull.}\ }\textbf {\bibinfo {volume} {63}},\ \bibinfo {pages} {336} (\bibinfo
  {year} {2018})}\BibitemShut {NoStop}%
\bibitem [{\citenamefont {Liu}\ \emph {et~al.}(2011)\citenamefont {Liu},
  \citenamefont {Huang}, \citenamefont {Li}, \citenamefont {Guo}, \citenamefont
  {Laine}, \citenamefont {Breuer},\ and\ \citenamefont {Piilo}}]{Liu2011}%
  \BibitemOpen
  \bibfield  {author} {\bibinfo {author} {\bibfnamefont {B.-H.}\ \bibnamefont
  {Liu}}, \bibinfo {author} {\bibfnamefont {Y.-F.}\ \bibnamefont {Huang}},
  \bibinfo {author} {\bibfnamefont {C.-F.}\ \bibnamefont {Li}}, \bibinfo
  {author} {\bibfnamefont {G.-C.}\ \bibnamefont {Guo}}, \bibinfo {author}
  {\bibfnamefont {E.-M.}\ \bibnamefont {Laine}}, \bibinfo {author}
  {\bibfnamefont {H.-P.}\ \bibnamefont {Breuer}}, \ and\ \bibinfo {author}
  {\bibfnamefont {J.}~\bibnamefont {Piilo}},\ }\href {\doibase
  10.1038/nphys2085} {\bibfield  {journal} {\bibinfo  {journal} {Nat. Phys.}\
  }\textbf {\bibinfo {volume} {7}},\ \bibinfo {pages} {931} (\bibinfo {year}
  {2011})}\BibitemShut {NoStop}%
\bibitem [{\citenamefont {Ho}\ \emph {et~al.}(2019)\citenamefont {Ho},
  \citenamefont {Matsuzaki}, \citenamefont {Matsuzaki},\ and\ \citenamefont
  {Kondo}}]{Ho2019}%
  \BibitemOpen
  \bibfield  {author} {\bibinfo {author} {\bibfnamefont {L.~B.}\ \bibnamefont
  {Ho}}, \bibinfo {author} {\bibfnamefont {Y.}~\bibnamefont {Matsuzaki}},
  \bibinfo {author} {\bibfnamefont {M.}~\bibnamefont {Matsuzaki}}, \ and\
  \bibinfo {author} {\bibfnamefont {Y.}~\bibnamefont {Kondo}},\ }\href
  {\doibase 10.1088/1367-2630/ab3a25} {\bibfield  {journal} {\bibinfo
  {journal} {New J. Phys.}\ }\textbf {\bibinfo {volume} {21}},\ \bibinfo
  {pages} {093008} (\bibinfo {year} {2019})}\BibitemShut {NoStop}%
\bibitem [{\citenamefont {Liu}\ \emph {et~al.}(2018)\citenamefont {Liu},
  \citenamefont {Lyyra}, \citenamefont {Sun}, \citenamefont {Liu},
  \citenamefont {Li}, \citenamefont {Guo}, \citenamefont {Maniscalco},\ and\
  \citenamefont {Piilo}}]{Liu2018}%
  \BibitemOpen
  \bibfield  {author} {\bibinfo {author} {\bibfnamefont {Z.-D.}\ \bibnamefont
  {Liu}}, \bibinfo {author} {\bibfnamefont {H.}~\bibnamefont {Lyyra}}, \bibinfo
  {author} {\bibfnamefont {Y.-N.}\ \bibnamefont {Sun}}, \bibinfo {author}
  {\bibfnamefont {B.-H.}\ \bibnamefont {Liu}}, \bibinfo {author} {\bibfnamefont
  {C.-F.}\ \bibnamefont {Li}}, \bibinfo {author} {\bibfnamefont {C.-G.}\
  \bibnamefont {Guo}}, \bibinfo {author} {\bibfnamefont {S.}~\bibnamefont
  {Maniscalco}}, \ and\ \bibinfo {author} {\bibfnamefont {S.}~\bibnamefont
  {Piilo}},\ }\href {\doibase 10.1038/s41467-018-05817-x} {\bibfield  {journal}
  {\bibinfo  {journal} {Nat. Comm.}\ }\textbf {\bibinfo {volume} {9}},\
  \bibinfo {pages} {3453} (\bibinfo {year} {2018})}\BibitemShut {NoStop}%
\bibitem [{\citenamefont {Bellomo}\ \emph {et~al.}(2007)\citenamefont
  {Bellomo}, \citenamefont {Lo~Franco},\ and\ \citenamefont
  {Compagno}}]{Bellomo2007}%
  \BibitemOpen
  \bibfield  {author} {\bibinfo {author} {\bibfnamefont {B.}~\bibnamefont
  {Bellomo}}, \bibinfo {author} {\bibfnamefont {R.}~\bibnamefont {Lo~Franco}},
  \ and\ \bibinfo {author} {\bibfnamefont {G.}~\bibnamefont {Compagno}},\
  }\href {\doibase 10.1103/PhysRevLett.99.160502} {\bibfield  {journal}
  {\bibinfo  {journal} {Phys. Rev. Lett.}\ }\textbf {\bibinfo {volume} {99}},\
  \bibinfo {pages} {160502} (\bibinfo {year} {2007})}\BibitemShut {NoStop}%
\bibitem [{\citenamefont {Bylicka}\ \emph {et~al.}(2014)\citenamefont
  {Bylicka}, \citenamefont {Chru{\'s}ci{\'n}ski},\ and\ \citenamefont
  {Maniscalco}}]{Bylicka2014}%
  \BibitemOpen
  \bibfield  {author} {\bibinfo {author} {\bibfnamefont {B.}~\bibnamefont
  {Bylicka}}, \bibinfo {author} {\bibfnamefont {D.}~\bibnamefont
  {Chru{\'s}ci{\'n}ski}}, \ and\ \bibinfo {author} {\bibfnamefont
  {S.}~\bibnamefont {Maniscalco}},\ }\href
  {http://dx.doi.org/10.1038/srep05720} {\bibfield  {journal} {\bibinfo
  {journal} {Sci. Rep.}\ }\textbf {\bibinfo {volume} {4}},\ \bibinfo {pages}
  {5720} (\bibinfo {year} {2014})}\BibitemShut {NoStop}%
\bibitem [{\citenamefont {Chin}\ \emph {et~al.}(2012)\citenamefont {Chin},
  \citenamefont {Huelga},\ and\ \citenamefont {Plenio}}]{Chin2012}%
  \BibitemOpen
  \bibfield  {author} {\bibinfo {author} {\bibfnamefont {A.~W.}\ \bibnamefont
  {Chin}}, \bibinfo {author} {\bibfnamefont {S.~F.}\ \bibnamefont {Huelga}}, \
  and\ \bibinfo {author} {\bibfnamefont {M.~B.}\ \bibnamefont {Plenio}},\
  }\href {\doibase 10.1103/PhysRevLett.109.233601} {\bibfield  {journal}
  {\bibinfo  {journal} {Phys. Rev. Lett.}\ }\textbf {\bibinfo {volume} {109}},\
  \bibinfo {pages} {233601} (\bibinfo {year} {2012})}\BibitemShut {NoStop}%
\bibitem [{\citenamefont {Maniscalco}\ \emph {et~al.}(2008)\citenamefont
  {Maniscalco}, \citenamefont {Francica}, \citenamefont {Zaffino},
  \citenamefont {Lo~Gullo},\ and\ \citenamefont {Plastina}}]{Maniscalco2008}%
  \BibitemOpen
  \bibfield  {author} {\bibinfo {author} {\bibfnamefont {S.}~\bibnamefont
  {Maniscalco}}, \bibinfo {author} {\bibfnamefont {F.}~\bibnamefont
  {Francica}}, \bibinfo {author} {\bibfnamefont {R.~L.}\ \bibnamefont
  {Zaffino}}, \bibinfo {author} {\bibfnamefont {N.}~\bibnamefont {Lo~Gullo}}, \
  and\ \bibinfo {author} {\bibfnamefont {F.}~\bibnamefont {Plastina}},\ }\href
  {\doibase 10.1103/PhysRevLett.100.090503} {\bibfield  {journal} {\bibinfo
  {journal} {Phys. Rev. Lett.}\ }\textbf {\bibinfo {volume} {100}},\ \bibinfo
  {pages} {090503} (\bibinfo {year} {2008})}\BibitemShut {NoStop}%
\bibitem [{\citenamefont {Gorini}\ \emph {et~al.}(1976)\citenamefont {Gorini},
  \citenamefont {Kossakowski},\ and\ \citenamefont {Sudarshan}}]{Gorini1976}%
  \BibitemOpen
  \bibfield  {author} {\bibinfo {author} {\bibfnamefont {V.}~\bibnamefont
  {Gorini}}, \bibinfo {author} {\bibfnamefont {A.}~\bibnamefont {Kossakowski}},
  \ and\ \bibinfo {author} {\bibfnamefont {E.~C.~G.}\ \bibnamefont
  {Sudarshan}},\ }\href {\doibase 10.1063/1.522979} {\bibfield  {journal}
  {\bibinfo  {journal} {J. Math. Phys.}\ }\textbf {\bibinfo {volume} {17}},\
  \bibinfo {pages} {821} (\bibinfo {year} {1976})}\BibitemShut {NoStop}%
\bibitem [{\citenamefont {Lindblad}(1976)}]{Lindblad1976}%
  \BibitemOpen
  \bibfield  {author} {\bibinfo {author} {\bibfnamefont {G.}~\bibnamefont
  {Lindblad}},\ }\href@noop {} {\bibfield  {journal} {\bibinfo  {journal}
  {Comm. Math. Phys.}\ }\textbf {\bibinfo {volume} {48}},\ \bibinfo {pages}
  {119} (\bibinfo {year} {1976})}\BibitemShut {NoStop}%
\bibitem [{\citenamefont {Dalibard}\ \emph {et~al.}(1992)\citenamefont
  {Dalibard}, \citenamefont {Castin},\ and\ \citenamefont
  {M\o{}lmer}}]{Dalibard1992}%
  \BibitemOpen
  \bibfield  {author} {\bibinfo {author} {\bibfnamefont {J.}~\bibnamefont
  {Dalibard}}, \bibinfo {author} {\bibfnamefont {Y.}~\bibnamefont {Castin}}, \
  and\ \bibinfo {author} {\bibfnamefont {K.}~\bibnamefont {M\o{}lmer}},\ }\href
  {\doibase 10.1103/PhysRevLett.68.580} {\bibfield  {journal} {\bibinfo
  {journal} {Phys. Rev. Lett.}\ }\textbf {\bibinfo {volume} {68}},\ \bibinfo
  {pages} {580} (\bibinfo {year} {1992})}\BibitemShut {NoStop}%
\bibitem [{\citenamefont {Plenio}\ and\ \citenamefont
  {Knight}(1998)}]{Plenio1998}%
  \BibitemOpen
  \bibfield  {author} {\bibinfo {author} {\bibfnamefont {M.~B.}\ \bibnamefont
  {Plenio}}\ and\ \bibinfo {author} {\bibfnamefont {P.~L.}\ \bibnamefont
  {Knight}},\ }\href {\doibase 10.1103/RevModPhys.70.101} {\bibfield  {journal}
  {\bibinfo  {journal} {Rev. Mod. Phys.}\ }\textbf {\bibinfo {volume} {70}},\
  \bibinfo {pages} {101} (\bibinfo {year} {1998})}\BibitemShut {NoStop}%
\bibitem [{\citenamefont {Gisin}\ and\ \citenamefont
  {Percival}(1992)}]{Gisin1992}%
  \BibitemOpen
  \bibfield  {author} {\bibinfo {author} {\bibfnamefont {N.}~\bibnamefont
  {Gisin}}\ and\ \bibinfo {author} {\bibfnamefont {I.~C.}\ \bibnamefont
  {Percival}},\ }\href {\doibase 10.1088/0305-4470/25/21/023} {\bibfield
  {journal} {\bibinfo  {journal} {J. Phys. A: Math. Gen.}\ }\textbf {\bibinfo
  {volume} {25}},\ \bibinfo {pages} {5677} (\bibinfo {year}
  {1992})}\BibitemShut {NoStop}%
\bibitem [{\citenamefont {Nakajima}(1958)}]{Nakajima1958}%
  \BibitemOpen
  \bibfield  {author} {\bibinfo {author} {\bibfnamefont {S.}~\bibnamefont
  {Nakajima}},\ }\href {\doibase 10.1143/PTP.20.948} {\bibfield  {journal}
  {\bibinfo  {journal} {Prog. Theor. Phys.}\ }\textbf {\bibinfo {volume}
  {20}},\ \bibinfo {pages} {948} (\bibinfo {year} {1958})}\BibitemShut
  {NoStop}%
\bibitem [{\citenamefont {Zwanzig}(1960)}]{Zwanzig1960}%
  \BibitemOpen
  \bibfield  {author} {\bibinfo {author} {\bibfnamefont {R.}~\bibnamefont
  {Zwanzig}},\ }\href {\doibase 10.1063/1.1731409} {\bibfield  {journal}
  {\bibinfo  {journal} {J. Chem. Phys.}\ }\textbf {\bibinfo {volume} {33}},\
  \bibinfo {pages} {1338} (\bibinfo {year} {1960})}\BibitemShut {NoStop}%
\bibitem [{\citenamefont {Imamoglu}(1994)}]{Imamoglu1994}%
  \BibitemOpen
  \bibfield  {author} {\bibinfo {author} {\bibfnamefont {A.}~\bibnamefont
  {Imamoglu}},\ }\href {\doibase 10.1103/PhysRevA.50.3650} {\bibfield
  {journal} {\bibinfo  {journal} {Phys. Rev. A}\ }\textbf {\bibinfo {volume}
  {50}},\ \bibinfo {pages} {3650} (\bibinfo {year} {1994})}\BibitemShut
  {NoStop}%
\bibitem [{\citenamefont {Stenius}\ and\ \citenamefont
  {Imamoglu}(1996)}]{Stenius1996}%
  \BibitemOpen
  \bibfield  {author} {\bibinfo {author} {\bibfnamefont {P.}~\bibnamefont
  {Stenius}}\ and\ \bibinfo {author} {\bibfnamefont {A.}~\bibnamefont
  {Imamoglu}},\ }\href {http://stacks.iop.org/1355-5111/8/i=1/a=021} {\bibfield
   {journal} {\bibinfo  {journal} {Quantum Semiclass. Opt.}\ }\textbf {\bibinfo
  {volume} {8}},\ \bibinfo {pages} {283} (\bibinfo {year} {1996})}\BibitemShut
  {NoStop}%
\bibitem [{\citenamefont {Chin}\ \emph {et~al.}(2010)\citenamefont {Chin},
  \citenamefont {Rivas}, \citenamefont {Huelga},\ and\ \citenamefont
  {Plenio}}]{Chin2010}%
  \BibitemOpen
  \bibfield  {author} {\bibinfo {author} {\bibfnamefont {A.~W.}\ \bibnamefont
  {Chin}}, \bibinfo {author} {\bibfnamefont {{\'A}.}~\bibnamefont {Rivas}},
  \bibinfo {author} {\bibfnamefont {S.~F.}\ \bibnamefont {Huelga}}, \ and\
  \bibinfo {author} {\bibfnamefont {M.~B.}\ \bibnamefont {Plenio}},\ }\href
  {\doibase 10.1063/1.3490188} {\bibfield  {journal} {\bibinfo  {journal} {J.
  Math. Phys.}\ }\textbf {\bibinfo {volume} {51}},\ \bibinfo {pages} {092109}
  (\bibinfo {year} {2010})}\BibitemShut {NoStop}%
\bibitem [{\citenamefont {Woods}\ \emph {et~al.}(2014)\citenamefont {Woods},
  \citenamefont {Groux}, \citenamefont {Chin}, \citenamefont {Huelga},\ and\
  \citenamefont {Plenio}}]{Woods2014}%
  \BibitemOpen
  \bibfield  {author} {\bibinfo {author} {\bibfnamefont {M.~P.}\ \bibnamefont
  {Woods}}, \bibinfo {author} {\bibfnamefont {R.}~\bibnamefont {Groux}},
  \bibinfo {author} {\bibfnamefont {A.~W.}\ \bibnamefont {Chin}}, \bibinfo
  {author} {\bibfnamefont {S.~F.}\ \bibnamefont {Huelga}}, \ and\ \bibinfo
  {author} {\bibfnamefont {M.~B.}\ \bibnamefont {Plenio}},\ }\href {\doibase
  10.1063/1.4866769} {\bibfield  {journal} {\bibinfo  {journal} {J. Math.
  Phys.}\ }\textbf {\bibinfo {volume} {55}},\ \bibinfo {pages} {032101}
  (\bibinfo {year} {2014})}\BibitemShut {NoStop}%
\bibitem [{\citenamefont {Tamascelli}\ \emph {et~al.}(2019)\citenamefont
  {Tamascelli}, \citenamefont {Smirne}, \citenamefont {Lim}, \citenamefont
  {Huelga},\ and\ \citenamefont {Plenio}}]{Tamascelli2019}%
  \BibitemOpen
  \bibfield  {author} {\bibinfo {author} {\bibfnamefont {D.}~\bibnamefont
  {Tamascelli}}, \bibinfo {author} {\bibfnamefont {A.}~\bibnamefont {Smirne}},
  \bibinfo {author} {\bibfnamefont {J.}~\bibnamefont {Lim}}, \bibinfo {author}
  {\bibfnamefont {S.~F.}\ \bibnamefont {Huelga}}, \ and\ \bibinfo {author}
  {\bibfnamefont {M.~B.}\ \bibnamefont {Plenio}},\ }\href {\doibase
  10.1103/PhysRevLett.123.090402} {\bibfield  {journal} {\bibinfo  {journal}
  {Phys. Rev. Lett.}\ }\textbf {\bibinfo {volume} {123}},\ \bibinfo {pages}
  {090402} (\bibinfo {year} {2019})}\BibitemShut {NoStop}%
\bibitem [{\citenamefont {Iles-Smith}\ \emph {et~al.}(2014)\citenamefont
  {Iles-Smith}, \citenamefont {Lambert},\ and\ \citenamefont
  {Nazir}}]{ISmith2014}%
  \BibitemOpen
  \bibfield  {author} {\bibinfo {author} {\bibfnamefont {J.}~\bibnamefont
  {Iles-Smith}}, \bibinfo {author} {\bibfnamefont {N.}~\bibnamefont {Lambert}},
  \ and\ \bibinfo {author} {\bibfnamefont {A.}~\bibnamefont {Nazir}},\ }\href
  {\doibase 10.1103/PhysRevA.90.032114} {\bibfield  {journal} {\bibinfo
  {journal} {Phys. Rev. A}\ }\textbf {\bibinfo {volume} {90}},\ \bibinfo
  {pages} {032114} (\bibinfo {year} {2014})}\BibitemShut {NoStop}%
\bibitem [{\citenamefont {Iles-Smith}\ \emph {et~al.}(2016)\citenamefont
  {Iles-Smith}, \citenamefont {Dijkstra}, \citenamefont {Lambert},\ and\
  \citenamefont {Nazir}}]{ISmith2016}%
  \BibitemOpen
  \bibfield  {author} {\bibinfo {author} {\bibfnamefont {J.}~\bibnamefont
  {Iles-Smith}}, \bibinfo {author} {\bibfnamefont {A.~G.}\ \bibnamefont
  {Dijkstra}}, \bibinfo {author} {\bibfnamefont {N.}~\bibnamefont {Lambert}}, \
  and\ \bibinfo {author} {\bibfnamefont {A.}~\bibnamefont {Nazir}},\ }\href
  {\doibase 10.1063/1.4940218} {\bibfield  {journal} {\bibinfo  {journal} {J.
  Chem. Phys.}\ }\textbf {\bibinfo {volume} {144}},\ \bibinfo {pages} {044110}
  (\bibinfo {year} {2016})}\BibitemShut {NoStop}%
\bibitem [{\citenamefont {Strasberg}\ \emph {et~al.}(2016)\citenamefont
  {Strasberg}, \citenamefont {Schaller}, \citenamefont {Lambert},\ and\
  \citenamefont {Brandes}}]{Strasberg2016}%
  \BibitemOpen
  \bibfield  {author} {\bibinfo {author} {\bibfnamefont {P.}~\bibnamefont
  {Strasberg}}, \bibinfo {author} {\bibfnamefont {G.}~\bibnamefont {Schaller}},
  \bibinfo {author} {\bibfnamefont {N.}~\bibnamefont {Lambert}}, \ and\
  \bibinfo {author} {\bibfnamefont {T.}~\bibnamefont {Brandes}},\ }\href
  {http://stacks.iop.org/1367-2630/18/i=7/a=073007} {\bibfield  {journal}
  {\bibinfo  {journal} {New J. Phys.}\ }\textbf {\bibinfo {volume} {18}},\
  \bibinfo {pages} {073007} (\bibinfo {year} {2016})}\BibitemShut {NoStop}%
\bibitem [{\citenamefont {Roden}\ \emph {et~al.}(2011)\citenamefont {Roden},
  \citenamefont {Strunz},\ and\ \citenamefont {Eisfeld}}]{Roden2011}%
  \BibitemOpen
  \bibfield  {author} {\bibinfo {author} {\bibfnamefont {J.}~\bibnamefont
  {Roden}}, \bibinfo {author} {\bibfnamefont {W.~T.}\ \bibnamefont {Strunz}}, \
  and\ \bibinfo {author} {\bibfnamefont {A.}~\bibnamefont {Eisfeld}},\ }\href
  {\doibase 10.1063/1.3512979} {\bibfield  {journal} {\bibinfo  {journal} {J.
  Chem. Phys.}\ }\textbf {\bibinfo {volume} {134}},\ \bibinfo {pages} {034902}
  (\bibinfo {year} {2011})}\BibitemShut {NoStop}%
\bibitem [{\citenamefont {Roden}\ \emph {et~al.}(2012)\citenamefont {Roden},
  \citenamefont {Strunz}, \citenamefont {Whaley},\ and\ \citenamefont
  {Eisfeld}}]{Roden2012}%
  \BibitemOpen
  \bibfield  {author} {\bibinfo {author} {\bibfnamefont {J.}~\bibnamefont
  {Roden}}, \bibinfo {author} {\bibfnamefont {W.~T.}\ \bibnamefont {Strunz}},
  \bibinfo {author} {\bibfnamefont {K.~B.}\ \bibnamefont {Whaley}}, \ and\
  \bibinfo {author} {\bibfnamefont {A.}~\bibnamefont {Eisfeld}},\ }\href
  {\doibase 10.1063/1.4765329} {\bibfield  {journal} {\bibinfo  {journal} {J.
  Chem. Phys.}\ }\textbf {\bibinfo {volume} {137}},\ \bibinfo {pages} {204110}
  (\bibinfo {year} {2012})}\BibitemShut {NoStop}%
\bibitem [{\citenamefont {Sch\"onleber}\ \emph {et~al.}(2015)\citenamefont
  {Sch\"onleber}, \citenamefont {Croy},\ and\ \citenamefont
  {Eisfeld}}]{Schonleber2015}%
  \BibitemOpen
  \bibfield  {author} {\bibinfo {author} {\bibfnamefont {D.~W.}\ \bibnamefont
  {Sch\"onleber}}, \bibinfo {author} {\bibfnamefont {A.}~\bibnamefont {Croy}},
  \ and\ \bibinfo {author} {\bibfnamefont {A.}~\bibnamefont {Eisfeld}},\ }\href
  {\doibase 10.1103/PhysRevA.91.052108} {\bibfield  {journal} {\bibinfo
  {journal} {Phys. Rev. A}\ }\textbf {\bibinfo {volume} {91}},\ \bibinfo
  {pages} {052108} (\bibinfo {year} {2015})}\BibitemShut {NoStop}%
\bibitem [{\citenamefont {Arrigoni}\ \emph {et~al.}(2013)\citenamefont
  {Arrigoni}, \citenamefont {Knap},\ and\ \citenamefont {von~der
  Linden}}]{Arrigoni2013}%
  \BibitemOpen
  \bibfield  {author} {\bibinfo {author} {\bibfnamefont {E.}~\bibnamefont
  {Arrigoni}}, \bibinfo {author} {\bibfnamefont {M.}~\bibnamefont {Knap}}, \
  and\ \bibinfo {author} {\bibfnamefont {W.}~\bibnamefont {von~der Linden}},\
  }\href {\doibase 10.1103/PhysRevLett.110.086403} {\bibfield  {journal}
  {\bibinfo  {journal} {Phys. Rev. Lett.}\ }\textbf {\bibinfo {volume} {110}},\
  \bibinfo {pages} {086403} (\bibinfo {year} {2013})}\BibitemShut {NoStop}%
\bibitem [{\citenamefont {Dorda}\ \emph {et~al.}(2014)\citenamefont {Dorda},
  \citenamefont {Nuss}, \citenamefont {von~der Linden},\ and\ \citenamefont
  {Arrigoni}}]{Dorda2013}%
  \BibitemOpen
  \bibfield  {author} {\bibinfo {author} {\bibfnamefont {A.}~\bibnamefont
  {Dorda}}, \bibinfo {author} {\bibfnamefont {M.}~\bibnamefont {Nuss}},
  \bibinfo {author} {\bibfnamefont {W.}~\bibnamefont {von~der Linden}}, \ and\
  \bibinfo {author} {\bibfnamefont {E.}~\bibnamefont {Arrigoni}},\ }\href
  {\doibase 10.1103/PhysRevB.89.165105} {\bibfield  {journal} {\bibinfo
  {journal} {Phys. Rev. B}\ }\textbf {\bibinfo {volume} {89}},\ \bibinfo
  {pages} {165105} (\bibinfo {year} {2014})}\BibitemShut {NoStop}%
\bibitem [{\citenamefont {Dorda}\ \emph {et~al.}(2017)\citenamefont {Dorda},
  \citenamefont {Sorantin}, \citenamefont {von~der Linden},\ and\ \citenamefont
  {Arrigoni}}]{Dorda2017}%
  \BibitemOpen
  \bibfield  {author} {\bibinfo {author} {\bibfnamefont {A.}~\bibnamefont
  {Dorda}}, \bibinfo {author} {\bibfnamefont {M.}~\bibnamefont {Sorantin}},
  \bibinfo {author} {\bibfnamefont {W.}~\bibnamefont {von~der Linden}}, \ and\
  \bibinfo {author} {\bibfnamefont {E.}~\bibnamefont {Arrigoni}},\ }\href
  {\doibase 10.1088/1367-2630/aa6ccc} {\bibfield  {journal} {\bibinfo
  {journal} {New J. Phys.}\ }\textbf {\bibinfo {volume} {19}},\ \bibinfo
  {pages} {063005} (\bibinfo {year} {2017})}\BibitemShut {NoStop}%
\bibitem [{\citenamefont {Pleasance}\ and\ \citenamefont
  {Garraway}(2017)}]{Pleasance2017}%
  \BibitemOpen
  \bibfield  {author} {\bibinfo {author} {\bibfnamefont {G.}~\bibnamefont
  {Pleasance}}\ and\ \bibinfo {author} {\bibfnamefont {B.~M.}\ \bibnamefont
  {Garraway}},\ }\href {\doibase 10.1103/PhysRevA.96.062105} {\bibfield
  {journal} {\bibinfo  {journal} {Phys. Rev. A}\ }\textbf {\bibinfo {volume}
  {96}},\ \bibinfo {pages} {062105} (\bibinfo {year} {2017})}\BibitemShut
  {NoStop}%
\bibitem [{\citenamefont {Garraway}(1997)}]{Garraway1997}%
  \BibitemOpen
  \bibfield  {author} {\bibinfo {author} {\bibfnamefont {B.~M.}\ \bibnamefont
  {Garraway}},\ }\href {\doibase 10.1103/PhysRevA.55.2290} {\bibfield
  {journal} {\bibinfo  {journal} {Phys. Rev. A.}\ }\textbf {\bibinfo {volume}
  {55}},\ \bibinfo {pages} {2290} (\bibinfo {year} {1997})}\BibitemShut
  {NoStop}%
\bibitem [{\citenamefont {Dalton}\ \emph {et~al.}(2001)\citenamefont {Dalton},
  \citenamefont {Barnett},\ and\ \citenamefont {Garraway}}]{Dalton2001}%
  \BibitemOpen
  \bibfield  {author} {\bibinfo {author} {\bibfnamefont {B.~J.}\ \bibnamefont
  {Dalton}}, \bibinfo {author} {\bibfnamefont {S.~M.}\ \bibnamefont {Barnett}},
  \ and\ \bibinfo {author} {\bibfnamefont {B.~M.}\ \bibnamefont {Garraway}},\
  }\href {\doibase 10.1103/PhysRevA.64.053813} {\bibfield  {journal} {\bibinfo
  {journal} {Phys. Rev. A}\ }\textbf {\bibinfo {volume} {64}},\ \bibinfo
  {pages} {053813} (\bibinfo {year} {2001})}\BibitemShut {NoStop}%
\bibitem [{\citenamefont {Dalton}\ and\ \citenamefont
  {Garraway}(2003)}]{Dalton2003}%
  \BibitemOpen
  \bibfield  {author} {\bibinfo {author} {\bibfnamefont {B.~J.}\ \bibnamefont
  {Dalton}}\ and\ \bibinfo {author} {\bibfnamefont {B.~M.}\ \bibnamefont
  {Garraway}},\ }\href {\doibase 10.1103/PhysRevA.68.033809} {\bibfield
  {journal} {\bibinfo  {journal} {Phys. Rev. A}\ }\textbf {\bibinfo {volume}
  {68}},\ \bibinfo {pages} {033809} (\bibinfo {year} {2003})}\BibitemShut
  {NoStop}%
\bibitem [{\citenamefont {Garraway}\ and\ \citenamefont
  {Dalton}(2006)}]{Garraway2006}%
  \BibitemOpen
  \bibfield  {author} {\bibinfo {author} {\bibfnamefont {B.~M.}\ \bibnamefont
  {Garraway}}\ and\ \bibinfo {author} {\bibfnamefont {B.~J.}\ \bibnamefont
  {Dalton}},\ }\href {http://stacks.iop.org/0953-4075/39/i=15/a=S21} {\bibfield
   {journal} {\bibinfo  {journal} {J. Phys. B: At. Mol. Opt. Phys.}\ }\textbf
  {\bibinfo {volume} {39}},\ \bibinfo {pages} {S767} (\bibinfo {year}
  {2006})}\BibitemShut {NoStop}%
\bibitem [{\citenamefont {Tamascelli}\ \emph {et~al.}(2018)\citenamefont
  {Tamascelli}, \citenamefont {Smirne}, \citenamefont {Huelga},\ and\
  \citenamefont {Plenio}}]{Tamascelli2018}%
  \BibitemOpen
  \bibfield  {author} {\bibinfo {author} {\bibfnamefont {D.}~\bibnamefont
  {Tamascelli}}, \bibinfo {author} {\bibfnamefont {A.}~\bibnamefont {Smirne}},
  \bibinfo {author} {\bibfnamefont {S.~F.}\ \bibnamefont {Huelga}}, \ and\
  \bibinfo {author} {\bibfnamefont {M.~B.}\ \bibnamefont {Plenio}},\ }\href
  {\doibase 10.1103/PhysRevLett.120.030402} {\bibfield  {journal} {\bibinfo
  {journal} {Phys. Rev. Lett.}\ }\textbf {\bibinfo {volume} {120}},\ \bibinfo
  {pages} {030402} (\bibinfo {year} {2018})}\BibitemShut {NoStop}%
\bibitem [{\citenamefont {Chen}\ \emph {et~al.}(2019)\citenamefont {Chen},
  \citenamefont {Arrigoni},\ and\ \citenamefont {Galperin}}]{Chen2019}%
  \BibitemOpen
  \bibfield  {author} {\bibinfo {author} {\bibfnamefont {F.}~\bibnamefont
  {Chen}}, \bibinfo {author} {\bibfnamefont {E.}~\bibnamefont {Arrigoni}}, \
  and\ \bibinfo {author} {\bibfnamefont {M.}~\bibnamefont {Galperin}},\ }\href
  {\doibase 10.1088/1367-2630/ab5ec5} {\bibfield  {journal} {\bibinfo
  {journal} {New J. Phys.}\ }\textbf {\bibinfo {volume} {21}},\ \bibinfo
  {pages} {123035} (\bibinfo {year} {2019})}\BibitemShut {NoStop}%
\bibitem [{\citenamefont {Lambert}\ \emph {et~al.}(2019)\citenamefont
  {Lambert}, \citenamefont {Ahmed}, \citenamefont {Cirio},\ and\ \citenamefont
  {Nori}}]{Lambert2019}%
  \BibitemOpen
  \bibfield  {author} {\bibinfo {author} {\bibfnamefont {N.}~\bibnamefont
  {Lambert}}, \bibinfo {author} {\bibfnamefont {S.}~\bibnamefont {Ahmed}},
  \bibinfo {author} {\bibfnamefont {M.}~\bibnamefont {Cirio}}, \ and\ \bibinfo
  {author} {\bibfnamefont {F.}~\bibnamefont {Nori}},\ }\href {\doibase
  10.1038/s41467-019-11656-1} {\bibfield  {journal} {\bibinfo  {journal} {Nat.
  Commun.}\ }\textbf {\bibinfo {volume} {10}},\ \bibinfo {pages} {3721}
  (\bibinfo {year} {2019})}\BibitemShut {NoStop}%
\bibitem [{\citenamefont {Gardiner}\ and\ \citenamefont
  {Collett}(1985)}]{Gardiner1985}%
  \BibitemOpen
  \bibfield  {author} {\bibinfo {author} {\bibfnamefont {C.~W.}\ \bibnamefont
  {Gardiner}}\ and\ \bibinfo {author} {\bibfnamefont {M.~J.}\ \bibnamefont
  {Collett}},\ }\href {\doibase 10.1103/PhysRevA.31.3761} {\bibfield  {journal}
  {\bibinfo  {journal} {Phys. Rev. A}\ }\textbf {\bibinfo {volume} {31}},\
  \bibinfo {pages} {3761} (\bibinfo {year} {1985})}\BibitemShut {NoStop}%
\bibitem [{\citenamefont {Di{\'o}si}(2012)}]{Diosi2012}%
  \BibitemOpen
  \bibfield  {author} {\bibinfo {author} {\bibfnamefont {L.}~\bibnamefont
  {Di{\'o}si}},\ }\href {\doibase 10.1103/PhysRevA.85.034101} {\bibfield
  {journal} {\bibinfo  {journal} {Phys. Rev. A}\ }\textbf {\bibinfo {volume}
  {85}},\ \bibinfo {pages} {034101} (\bibinfo {year} {2012})}\BibitemShut
  {NoStop}%
\bibitem [{\citenamefont {Budini}(2013)}]{Budini2013}%
  \BibitemOpen
  \bibfield  {author} {\bibinfo {author} {\bibfnamefont {A.~A.}\ \bibnamefont
  {Budini}},\ }\href {\doibase 10.1103/PhysRevA.88.012124} {\bibfield
  {journal} {\bibinfo  {journal} {Phys. Rev. A}\ }\textbf {\bibinfo {volume}
  {88}},\ \bibinfo {pages} {012124} (\bibinfo {year} {2013})}\BibitemShut
  {NoStop}%
\bibitem [{\citenamefont {Breuer}(2004)}]{Breuer2004}%
  \BibitemOpen
  \bibfield  {author} {\bibinfo {author} {\bibfnamefont {H.-P.}\ \bibnamefont
  {Breuer}},\ }\href {\doibase 10.1103/PhysRevA.70.012106} {\bibfield
  {journal} {\bibinfo  {journal} {Phys. Rev. A}\ }\textbf {\bibinfo {volume}
  {70}},\ \bibinfo {pages} {012106} (\bibinfo {year} {2004})}\BibitemShut
  {NoStop}%
\bibitem [{\citenamefont {Mazzola}\ \emph {et~al.}(2009)\citenamefont
  {Mazzola}, \citenamefont {Maniscalco}, \citenamefont {Piilo}, \citenamefont
  {Suominen},\ and\ \citenamefont {Garraway}}]{Mazzola2009}%
  \BibitemOpen
  \bibfield  {author} {\bibinfo {author} {\bibfnamefont {L.}~\bibnamefont
  {Mazzola}}, \bibinfo {author} {\bibfnamefont {S.}~\bibnamefont {Maniscalco}},
  \bibinfo {author} {\bibfnamefont {J.}~\bibnamefont {Piilo}}, \bibinfo
  {author} {\bibfnamefont {K.-A.}\ \bibnamefont {Suominen}}, \ and\ \bibinfo
  {author} {\bibfnamefont {B.~M.}\ \bibnamefont {Garraway}},\ }\href {\doibase
  10.1103/PhysRevA.80.012104} {\bibfield  {journal} {\bibinfo  {journal} {Phys.
  Rev. A}\ }\textbf {\bibinfo {volume} {80}},\ \bibinfo {pages} {012104}
  (\bibinfo {year} {2009})}\BibitemShut {NoStop}%
\bibitem [{\citenamefont {Xu}\ \emph {et~al.}(2019)\citenamefont {Xu},
  \citenamefont {Thingna}, \citenamefont {Guo},\ and\ \citenamefont
  {Poletti}}]{Xu2019}%
  \BibitemOpen
  \bibfield  {author} {\bibinfo {author} {\bibfnamefont {X.}~\bibnamefont
  {Xu}}, \bibinfo {author} {\bibfnamefont {J.}~\bibnamefont {Thingna}},
  \bibinfo {author} {\bibfnamefont {C.}~\bibnamefont {Guo}}, \ and\ \bibinfo
  {author} {\bibfnamefont {D.}~\bibnamefont {Poletti}},\ }\href {\doibase
  10.1103/PhysRevA.99.012106} {\bibfield  {journal} {\bibinfo  {journal} {Phys.
  Rev. A}\ }\textbf {\bibinfo {volume} {99}},\ \bibinfo {pages} {012106}
  (\bibinfo {year} {2019})}\BibitemShut {NoStop}%
\bibitem [{\citenamefont {Daley}(2014)}]{Daley2014}%
  \BibitemOpen
  \bibfield  {author} {\bibinfo {author} {\bibfnamefont {A.~J.}\ \bibnamefont
  {Daley}},\ }\href {\doibase 10.1080/00018732.2014.933502} {\bibfield
  {journal} {\bibinfo  {journal} {Adv. Phys.}\ }\textbf {\bibinfo {volume}
  {63}},\ \bibinfo {pages} {77} (\bibinfo {year} {2014})}\BibitemShut {NoStop}%
\bibitem [{\citenamefont {Mascherpa}\ \emph {et~al.}(2020)\citenamefont
  {Mascherpa}, \citenamefont {Smirne}, \citenamefont {Somoza}, \citenamefont
  {Fern\'andez-Acebal}, \citenamefont {Donadi}, \citenamefont {Tamascelli},
  \citenamefont {Huelga},\ and\ \citenamefont {Plenio}}]{Mascherpa2019}%
  \BibitemOpen
  \bibfield  {author} {\bibinfo {author} {\bibfnamefont {F.}~\bibnamefont
  {Mascherpa}}, \bibinfo {author} {\bibfnamefont {A.}~\bibnamefont {Smirne}},
  \bibinfo {author} {\bibfnamefont {A.~D.}\ \bibnamefont {Somoza}}, \bibinfo
  {author} {\bibfnamefont {P.}~\bibnamefont {Fern\'andez-Acebal}}, \bibinfo
  {author} {\bibfnamefont {S.}~\bibnamefont {Donadi}}, \bibinfo {author}
  {\bibfnamefont {D.}~\bibnamefont {Tamascelli}}, \bibinfo {author}
  {\bibfnamefont {S.~F.}\ \bibnamefont {Huelga}}, \ and\ \bibinfo {author}
  {\bibfnamefont {M.~B.}\ \bibnamefont {Plenio}},\ }\href {\doibase
  10.1103/PhysRevA.101.052108} {\bibfield  {journal} {\bibinfo  {journal}
  {Phys. Rev. A}\ }\textbf {\bibinfo {volume} {101}},\ \bibinfo {pages}
  {052108} (\bibinfo {year} {2020})}\BibitemShut {NoStop}%
\bibitem [{\citenamefont {Chicone}(2006)}]{Chicone2006}%
  \BibitemOpen
  \bibfield  {author} {\bibinfo {author} {\bibfnamefont {C.}~\bibnamefont
  {Chicone}},\ }\href@noop {} {\emph {\bibinfo {title} {Ordinary Differential
  Equations with Applications}}}\ (\bibinfo  {publisher} {Springer},\ \bibinfo
  {address} {New York},\ \bibinfo {year} {2006})\BibitemShut {NoStop}%
\end{thebibliography}%

\end{document}